\documentclass[article,nojss]{jss}

\usepackage{orcidlink,thumbpdf,lmodern}

\usepackage{framed}
\usepackage{amsmath}
\usepackage{booktabs}
\usepackage{bbm}
\usepackage{pdflscape}
\usepackage{makecell}

\newcommand{\fct}[1]{\code{#1()}}

\author{Corissa T. Rohloff~\orcidlink{0000-0003-3228-4653}\\Human Resources Research Organization\\ (HumRRO)
   \And Rik Lamm~\orcidlink{0000-0002-3317-6243}\\Bloomington Public Schools
   \AND Yadira Peralta~\orcidlink{0000-0003-4823-6939}\\Center for Research and\\Teaching in Economics
   \And Nidhi Kohli~\orcidlink{0000-0003-4690-2854}\\University of Minnesota
   \And Eric F. Lock~\orcidlink{0000-0003-4663-2356}\\University of Minnesota}
\Plainauthor{Corissa T. Rohloff, Rik Lamm, Yadira Peralta, Nidhi Kohli, Eric F. Lock}

\title{\pkg{BEND}: An \proglang{R} Package for the Bayesian Estimation of Nonlinear Longitudinal Data}
\Plaintitle{BEND: An R Package for the Bayesian Estimation of Nonlinear Longitudinal Data}
\Shorttitle{BEND: Bayesian Estimation of Nonlinear Longitudinal Data}

\Abstract{
    Longitudinal data are useful for capturing and analyzing patterns of change over time. Often, these patterns follow a nonlinear form. One useful and commonly applied nonlinear function is the piecewise function, which assumes growth occurs in distinct phases, each with its own functional form. Past literature has established that Bayesian inference is preferred over likelihood-based methods for estimating piecewise models. To address this, we developed the \proglang{R} package \pkg{BEND} – Bayesian Estimation of Nonlinear Data (available on CRAN). The purpose of \pkg{BEND} is to provide a user-friendly software for estimating nonlinear longitudinal models using a Bayesian inference approach. Given the flexibility and practicality of the piecewise models, \pkg{BEND} includes several extensions of it to accommodate various types of complex longitudinal datasets and applications. \fct{Bayes\_PREM} can empirically identify the number and location of random changepoints in a piecewise random effects model. This function can also model multiple latent classes with different longitudinal growth patterns and incorporate covariates to predict the outcome and latent class membership. \fct{Bayes\_BPREM} can jointly model the longitudinal piecewise trajectories of two interrelated outcomes. Lastly, \fct{Bayes\_CREM} can estimate the impact of group membership on longitudinal growth. This paper provides an overview of the functions included in \pkg{BEND} and empirical examples of how to apply these models in practice.
}

\Plainkeywords{longitudinal methods, Bayesian inference, nonlinear forms, piecewise function}

\Address{
  Corissa T. Rohloff\\
  Human Resources Research Organization (HumRRO)\\
  100 Washington Avenue South\\
  Suite 1660\\
  Minneapolis, MN 55401\\
  E-mail: \email{crohloff@humrro.org}
}

\begin{document}

%% -- Introduction -------------------------------------------------------------

%% - In principle "as usual".
%% - But should typically have some discussion of both _software_ and _methods_.
%% - Use \proglang{}, \pkg{}, and \code{} markup throughout the manuscript.
%% - If such markup is in (sub)section titles, a plain text version has to be
%%   added as well.
%% - All software mentioned should be properly \cite-d.
%% - All abbreviations should be introduced.
%% - Unless the expansions of abbreviations are proper names (like "Journal
%%   of Statistical Software" above) they should be in sentence case (like
%%   "generalized linear models" below).

\section{Introduction} \label{sec:intro}

Longitudinal data are highly valuable across many disciplines for capturing and analyzing patterns of change over time \citep{Fitzmaurice2011Long}. Often, these patterns do not follow a constant rate of change, meaning they follow some nonlinear form. Nonlinear functional forms, especially intrinsically nonlinear forms, can be computationally challenging and difficult to reliably and accurately estimate \citep[e.g., piecewise models;][]{Kohli2016finite, Kohli2013Modeling}. Past literature has established that Bayesian inference is a suitable method for estimating these types of nonlinear models \citep{Kohli2015Fitting, Wang2008Simulation}. \pkg{BEND} is a package in the \proglang{R} environment that was developed for the Bayesian Estimation of Nonlinear Data \citep[BEND;][]{R}. The goal of this package is to provide a user-friendly software tool for estimating nonlinear longitudinal models for a variety of complex data structures using Bayesian inference approach, as well as tools to visualize and summarize the results. This package has been published and is freely available on the Comprehensive R Archive Network (CRAN): \hyperlink{https://CRAN.R-project.org/package=BEND}{https://CRAN.R-project.org/package=BEND}.

Longitudinal datasets are those where repeated measurements are collected on an outcome or variable of interest (e.g., achievement scores, body weight, reaction time) over time. These repeated measurements can be useful for visualizing and estimating growth (or loss) patterns over time. A popular set of models used to examine this type of data is called random effects models \citep[REMs;][]{Laird1982REMs}. REMs are well-suited for longitudinal data because they can account for the “nesting” embedded with this data structure (i.e., repeated measurements nested within individuals). Here is the equation for a basic linear REM,
\begin{equation} \label{eq:lrem}
Y_{ji} = \beta_{0i} + \beta_{1i}x_{ji} + \epsilon_{ji}
\end{equation}
where $Y_{ji}$ is the outcome at measurement occasion $j$ for individual $i$, $x_{ji}$ denotes the time variable at measurement occasion $j$ for individual $i$, and $\epsilon_{ji}$ is the error at measurement occasion $j$ for individual $i$. Both $\beta_{0i}$ and $\beta_{1i}$ are called random coefficients (for the intercept and slope, respectively) and can be decomposed as follows:
\begin{equation} \label{eq:randcoeff}
\beta_{qi} = \beta_{q} + b_{qi}
\end{equation}
Here, $\beta_{qi}$ is the random coefficient of the $q$th parameter for the $i$th individual. $\beta_{q}$ is the fixed or mean effect of the $q$th parameter, and $b_{qi}$ is the individual random effect associated with the $q$th parameter (i.e., individual deviation from the fixed effect). It is assumed that the random coefficients follow a multivariate normal distribution centered at the fixed effect with a variance-covariance matrix $\Phi_b$. This variance-covariance matrix accounts for the nested structure in longitudinal data and represents the amount of variation in the individual growth trajectories. 

As previously mentioned, the pattern of growth rarely follows a constant rate of change. Therefore, the linear REM equation can be modified to allow for nonlinear growth \citep{Lindstrom1990Nonlinear},
\begin{equation} \label{eq:nlrem}
Y_{ji} = f(\boldsymbol{\beta}_i,{x}_{ji}) + \epsilon_{ji}
\end{equation}
where $f$ represents a nonlinear functional form. We break down nonlinear functional forms into two categories: intrinsically linear and intrinsically nonlinear. Intrinsically nonlinear forms are defined as equations where at least one of the derivatives with respect to each parameter in the model will rely on at least one other model parameter \citep{Bates1988Nonlinear}. For example, an exponential model (denoted as, $a+b(1-e^{-cx})$) is intrinsically nonlinear because the derivatives will always rely on the rate parameter ($c$). An example of intrinsically linear forms that follow a nonlinear shape would be polynomials (i.e., quadratic, cubic) because one can always derive out the polynomial exponent. Intrinsically nonlinear forms are more difficult to estimate but are beneficial because they incorporate parameters that are easier to interpret from underlying substantive theory. With respect to the exponential model, regression coefficients associated with a total growth ($b$) and the growth rate parameter are likely more meaningful to the substantive researchers than that associated with a quadratic slope.

One useful and commonly applied intrinsically nonlinear form is called the piecewise function with unknown changepoint(s), 
\begin{equation} \label{eq:baseprem}
\begin{aligned} 
Y_{ji} = \beta_{0i} + \beta_{1i}x_{ji} + \beta_{2i}(x_{ji}-\gamma_i)^+ + \epsilon_{ji} \\
(x_{ji}-\gamma_i)^+ = 
    \begin{cases}
      x_{ji}-\gamma_i & \text{if $x_{ji}-\gamma_i>0$}\\
      0 & \text{otherwise}\\
    \end{cases}  
\end{aligned}
\end{equation}
and $\beta_{0i}$ is the random coefficient for the intercept, $\beta_{1i}$ is the random coefficient for the first slope, $\beta_{2i}$ is the random coefficient for the change in slope after the changepoint, $\gamma_i$ is the random coefficient for changepoint. A piecewise model assumes that underlying growth phenomena occurs in “phases.” The start and end of each phase is defined by the changepoint parameter which, when unknown and estimated empirically, is the intrinsically nonlinear parameter. 

The piecewise model is well-suited for nonlinear data because each growth phase can take on any functional form \citep{Cudeck2002RepMes}. Therefore, it has been used in many applications, such as modeling reading development \citep{Sullivan2017Longitudinal} and timing interventions \citep{Kreisman2003Evaluating}. In this manuscript, we focus on a linear-linear piecewise model. Due to the flexibility and practicality of the piecewise REM for longitudinal data, we have developed several extensions to accommodate various types of longitudinal datasets and applications. These extensions are all included as functions in the \proglang{R} package, \pkg{BEND}, which we will review here. Moreover, Bayesian estimation of piecewise models is preferred over likelihood-based estimation methods \citep{Kohli2015Fitting, Wang2008Simulation}. Thus, having a package that condenses several efforts to estimate Bayesian piecewise models is noteworthy. 

First, \cite{Lock2018Detecting} introduced a Bayesian piecewise REM that allows for multiple random changepoints where the number of changepoints can be empirically identified from the dataset. This model is useful in applications where the number and location of the changepoints are unknown and cannot be specified a priori. In addition to the empirical detection of multiple random changepoints, the authors allowed for empirical estimation of multiple latent (or unobserved) classes, also known as a mixture model. Therefore, the extensions in \cite{Lock2018Detecting} enable users to estimate the number of latent classes and the number and location of changepoints within each class. Subsequently, to further increase the utility of the piecewise REM, \cite{Lamm2022Incorporation} incorporated covariates to predict the longitudinal outcome and class membership. Adding these covariates allows researchers to explain variation in the outcome and the likelihood of belonging to one latent class vs. another. Each of these extensions is included in the \fct{Bayes\_PREM} function. 

In some substantive applications, two variables of interest may influence each other over time (e.g., reading and math achievement, the care a child receives and physical growth). That is, their development is intertwined, preventing the growth of each variable of interest in isolation. Under these scenarios, accounting for the joint multivariate nature of the data is relevant. To model these two correlated outcomes over time requires a bivariate modeling approach, called a bivariate REM model. \cite{Peralta2022Bayesian} developed a bivariate piecewise REM that allows for the joint estimation of two longitudinal outcomes. This extension is included in the \fct{Bayes\_BPREM} function. 

Lastly, in longitudinal data, the nesting structure maybe more complex than the typical structure of time nested within individuals. To elaborate, in many substantive applications, data structure may include additional layers of nesting beyond the typical time-within-individuals structure. For instance, individuals may be further nested within higher-level groups (e.g., students within schools or classrooms, patients within healthcare providers or hospitals). Furthermore, the group an individual belongs to may change over time. For example, students can transfer schools or patients can see different nurses throughout their care. A crossed REM allows us to evaluate the impact of this dynamic group membership on individual growth. \cite{Rohloff2024Identifiability} developed a piecewise crossed REM for longitudinal data along with other linear and nonlinear growth forms. This model is included in the \fct{Bayes\_CREM} function. 

For more information on the motivation, development, and performance of these piecewise REM extensions, we encourage readers to see the original papers for each model. Here, we will only provide the details necessary to use the \pkg{BEND} \proglang{R} package. In the following sections, we will describe the estimation framework used in the current \proglang{R} package, outline the modeling specifications for each function, provide examples of how to implement these models, and note some limitations of the current package. 

%% -- Bayesian Inference ---------------------------------------------------------------

\section{Bayesian Inference} \label{sec:bayes}

Each of the extensions introduced above comes with increased computational challenges and demands. Commonly used software programs currently implement a frequentist framework, using maximum likelihood (ML) estimation, to estimate REMs (e.g., \fct{MIXED}/\fct{NLMIXED} procedures in \proglang{SAS}, \pkg{nlme} and \pkg{lme4} packages in \proglang{R}). However, previous literature has demonstrated the advantage of using Bayesian inference for piecewise functional forms over the ML estimation approach \citep{Kohli2015Fitting}, mixture models \citep{Lamm2022Incorporation, Lock2018Detecting}, bivariate outcomes \citep{Peralta2022Bayesian}, and crossed random effects models \citep{Rohloff2024Identifiability} as it provides the best balance between accuracy, reliability, and computational efficiency. Many Bayesian software programs can estimate the various extensions of the piecewise REM (e.g., \proglang{STAN}, \proglang{JAGS}, \proglang{WINBUGS}). However, these programs require advanced knowledge of the statistical models and software. The \pkg{BEND} package provides a user-friendly method for fitting complex longitudinal models to nonlinear data using Bayesian inference. 

Rather than analytically deriving the likelihood or posterior distributions, which is often impossible for these high-dimensional models, Bayesian methods rely on sampling from various posterior distributions. The current \proglang{R} package will implement Gibbs sampling, a Markov chain Monte Carlo (MCMC) method, via the \pkg{rjags} package in \proglang{R} \citep{rjags}. The \pkg{rjags} package utilizes Just Another Gibbs Sampler (\proglang{JAGS}) software which implements Gibbs sampling to approximate the posterior distribution for each parameter \citep{plummer_jags_2003}. While other Bayesian inference software may be used, we choose JAGS for its efficiency and flexibility when sampling discrete parameters (e.g., latent class membership or changepoint indicators).   Typically, posterior sampling requires a set of throw-away iterations, called a “burn-in,” before taking a final set of iterations for inference. Here, we take the mean of the posterior samples as a point estimate of the posterior distribution for each parameter. In the following section, we will specify the default number of iterations used for burn-in and posterior sampling for each function in the \pkg{BEND} package. 

Lastly, to monitor convergence within the Bayesian framework we use the potential scale reduction factor (PSRF). The PSRF is a measure of the level of agreement between the MCMC chains that are run in parallel \citep{Brooks1998Convg, Gelman1992Inference}. In the current package, we will use three MCMC chains. The PSRF is provided for each parameter in the model. The multivariate PSRF was developed as an overall measure of model convergence \citep{Brooks1998Convg}. Typically, 1.1 and 1.2 are used as cutoffs \citep{Brooks1998Convg, Sinharay2004Convg}. However, with these highly parameterized models, the multivariate PSRF may be too conservative for assessing convergence. Thus, we also include the mean PSRF, the mean of all parameter PSRF values, to provide a less conservative measure of model convergence \citep{Lock2018Detecting, Peralta2022Bayesian, Rohloff2024Identifiability}.

%% -- Model Specification ---------------------------------------------------------------

\section{Model Specification} \label{sec:model}

This section will give an overview of the primary modeling functions in the \pkg{BEND} package. The prior distributions for each function are provided in the Appendices.

\subsection{BayesPREM}
  
The \fct{Bayes\_PREM} function includes the models developed by \cite{Lock2018Detecting} and \cite{Lamm2022Incorporation}: piecewise random effects model (PREM; default), covariate-influenced piecewise random effects model (CI-PREM), piecewise random effects mixture model (PREMM), and covariate-influenced piecewise random effects mixture model (CI-PREMM). Table~\ref{tab:latent_classes} displays a chart to help users decide which model is needed.

\begin{table}[h]
    \centering
    \begin{tabular}{lcc}
        \toprule
        & \multicolumn{2}{c}{What is the hypothesized number of latent classes?} \\
        \cmidrule(lr){2-3}
        Are covariates included? & 1 & 2+ \\
        \midrule
        No & PREM & PREMM \\
        Yes & CI-PREM & CI-PREMM \\
        \bottomrule
    \end{tabular}
    \caption{Models Included in the Bayes\_PREM Function}
    \label{tab:latent_classes}
\end{table}

\subsubsection{PREM}

The PREM allows users to estimate a piecewise growth trajectory when only one hypothesized latent class exists. This version of the PREM differs from Equation \ref{eq:baseprem} because the number and location of the changepoint parameters are estimated empirically. See the following equation,
\begin{equation} \label{eq:prem}
\begin{aligned} 
Y_{ji} = \beta_{0i} + \beta_{1i}x_{ji} +& \sum_{k=1}^{K} \beta_{\{k+1\}i}(x_{ji}-\gamma_{ki})^+ \mathbbm{1}_{\{k \leq \mathcal{K}\}} + \epsilon_{ji} \\
(x_{ji}-\gamma_{ki})^+ &= 
    \begin{cases}
      x_{ji}-\gamma_{ki} & \text{if $x_{ji}-\gamma_{ki}>0$}\\
      0 & \text{otherwise}\\
    \end{cases} \\
\mathbbm{1}_{\{k \leq \mathcal{K}\}} &=
    \begin{cases}
      1 & \text{if $k \leq \mathcal{K}$}\\
      0 & \text{otherwise}\\
    \end{cases}
\end{aligned}
\end{equation}
Here, $Y_{ji}$, $x_{ji}$, and $\epsilon_{ji}$ are defined as above; $\mathcal{K}\in{\{0,1,...,K\}}$ is a latent parameter indicating the unknown number of changepoints; $\beta_{0i}$, $\beta_{1i}$, $\beta_{\{k+1\}i}$, and $\gamma_{ki}$ are the intercept, first slope, change in slope after the $k$th changepoint, and the changepoint location for $k$th changepoint, respectively. The distributional assumptions and priors for the PREM are provided in Appendix~\ref{app:prem_app}. The PREM is the default model in the \fct{Bayes\_PREM} function, each of the following models builds off Equation \ref{eq:prem}. 

\subsubsection{CI-PREM}

The CI-PREM adds to the PREM by incorporating covariates to predict the outcome $Y_{ji}$. 
\begin{equation} \label{eq:ciprem}
\begin{aligned} 
Y_{ji} = \beta_{0i} + \beta_{1i}x_{ji} + \sum_{k=1}^{K} &\beta_{\{k+1\}i}(x_{ji}-\gamma_{ki})^+ \mathbbm{1}_{\{k \leq \mathcal{K}\}} + \Biggl\{ \sum_{p=1}^{P} \alpha_p w_{pi} \Biggl\} + \epsilon_{ji} \\
(x_{ji}-\gamma_{ki})^+ &= 
    \begin{cases}
      x_{ji}-\gamma_{ki} & \text{if $x_{ji}-\gamma_{ki}>0$}\\
      0 & \text{otherwise}\\
    \end{cases} \\
\mathbbm{1}_{\{k \leq \mathcal{K}\}} &=
    \begin{cases}
      1 & \text{if $k \leq \mathcal{K}$}\\
      0 & \text{otherwise}\\
    \end{cases}
\end{aligned}
\end{equation}
$w_{pi}$ is the observed data for outcome predictive covariate $p$ for individual $i$ (time-varying or -invariant) and $\alpha_p$ is the coefficient associated with outcome predictive covariate $p$. The distributional assumptions and priors for the CI-PREM are provided in Appendix~\ref{app:ciprem_app}.

\subsubsection{PREMM}

The PREMM expands upon the PREM by allowing the model to estimate the piecewise growth trajectory and the number/location of the changepoints for multiple latent classes.
\begin{equation} \label{eq:premm}
\begin{aligned} 
Y_{ji} = \beta_{0i} + \beta_{1i}x_{ji} + \sum_{k=1}^{K} &\beta_{\{k+1\}i}(x_{ji}-\gamma_{ki})^+ \mathbbm{1}_{\{k \leq \mathcal{K}_{\psi(i)}\}} + \epsilon_{ji} \\
(x_{ji}-\gamma_{ki})^+ &= 
    \begin{cases}
      x_{ji}-\gamma_{ki} & \text{if $x_{ji}-\gamma_{ki}>0$}\\
      0 & \text{otherwise}\\
    \end{cases} \\
\mathbbm{1}_{\{k \leq \mathcal{K}_{\psi(i)}\}} &=
    \begin{cases}
      1 & \text{if $k \leq \mathcal{K}_{\psi(i)}$}\\
      0 & \text{otherwise}\\
    \end{cases}
\end{aligned}
\end{equation}
$\psi(i)\in{\{1,...,C\}}$ is the class to which individual $i$ belongs (where $p(\psi(i)=c)=\pi_{ic}$ is the probability that individual $i$ belongs to class $c$), $\mathcal{K}\in{\{0,1,...,K\}}$ is the latent parameter indicating the unknown number of changepoints, and $\mathcal{K}_{\psi(i)}$  is the latent number of changepoints in each class. Beyond the number of changepoints, the fixed effects for each parameter also depend on class membership $\psi(i)$; the distributional assumptions and priors for the PREMM are provided in Appendix~\ref{app:premm_app}.

\subsubsection{CI-PREMM}

The CI-PREMM builds off the PREMM in Equation \ref{eq:premm} by incorporating covariates to predict the outcome $Y_{ji}$ and class membership. This model can be broken down into two equations. The first equation predicts class membership using a logistic regression model.

\begin{equation} \label{eq:cipremm1}
\log \Bigl( \frac{\pi_{ic}}{\pi_{iC}}\Bigl) = \log \Bigl( \frac{\pi_{ic}}{1-\text{$\sum_{c=1}^{C-1}\pi_{ic}$}}\Bigl) = \sum_{l=0}^{L}\lambda_{cl}z_{li} \\
\end{equation}

Here, let $c\in{\{1,...,C-1\}}$ due to one of the classes being the reference class. Because class membership is predicted using logistic regression, only two latent classes ($C=2$) can be hypothesize for this model. $\pi_{ic}$ is the probability that individual $i$ belongs to class $c$ and $\pi_{iC}$ is the probability that individual $i$ belongs to reference class ($c=1$). $z_{li}$ is the observed data for class predictive covariate $l$ for individual $i$ which must be time-invariant. $z_0=1$ for every class to allow $\lambda_{0c}$ to be the intercept for each of the $C-1$ classes. Lastly, $\lambda_{cl}$ is the coefficient associated with class predictive covariate $l$ for class $c$. Because we are only hypothesizing two latent classes in the CI-PREMM, there is only one set of regression estimates for the class-predictive covariates ($c=2$). 

The next equation in the CI-PREMM incorporates time-invariant and/or time-varying covariates to predict the outcome as was done in CI-PREM in Equation \ref{eq:ciprem}. 

\begin{equation} \label{eq:cipremm2}
\begin{aligned} 
Y_{ji} = \beta_{0i} + \beta_{1i}x_{ji} + \sum_{k=1}^{K} &\beta_{\{k+1\}i}(x_{ji}-\gamma_{ki})^+ \mathbbm{1}_{\{k \leq \mathcal{K}_{\psi(i)}\}} + \Biggl\{ \sum_{p=1}^{P} \alpha_p w_{pi} \Biggl\} + \epsilon_{ji} \\
(x_{ji}-\gamma_{ki})^+ &= 
    \begin{cases}
      x_{ji}-\gamma_{ki} & \text{if $x_{ji}-\gamma_{ki}>0$}\\
      0 & \text{otherwise}\\
    \end{cases} \\
\mathbbm{1}_{\{k \leq \mathcal{K}_{\psi(i)}\}} &=
    \begin{cases}
      1 & \text{if $k \leq \mathcal{K}_{\psi(i)}$}\\
      0 & \text{otherwise}\\
    \end{cases}
\end{aligned}
\end{equation}

Note, $\alpha_p$ does not have a subscript $c$. Thus, the coefficient associated with outcome predictive covariate $p$ is assumed to be constant across latent classes. The distributional assumptions and priors for the CI-PREMM are provided in Appendix~\ref{app:cipremm_app}.

\subsection{BayesBPREM}

The \fct{Bayes\_BPREM} function fits a bivariate piecewise random effects model (BPREM) developed by \cite{Peralta2022Bayesian}. This allows for the joint modeling of two related outcomes ($Y_{1ji}$ and $Y_{2ji}$), as follows: 

\begin{equation} \label{eq:bprem}
\begin{aligned} 
\begin{cases}
   Y_{1ji} = \beta_{10i} + \beta_{11i}x_{ji} + \beta_{12i}(x_{ji} - \gamma_{1i})^+ + \epsilon_{1ji} \\
   Y_{2ji} = \beta_{20i} + \beta_{21i}x_{ji} + \beta_{22i}(x_{ji} - \gamma_{2i})^+ + \epsilon_{2ji} \\
\end{cases} \\
(x_{ji}-\gamma_{\cdot i})^+ = 
    \begin{cases}
      x_{ji}-\gamma_{\cdot i} & \text{if $x_{ji}-\gamma_{\cdot i} > 0$}\\
      0 & \text{otherwise}\\
    \end{cases} \\
\end{aligned}
\end{equation}

For each outcome, where $\cdot$ in each parameter denotes outcomes 1 and 2, $\beta_{\cdot 0i}$, $\beta_{\cdot 1i}$, $\beta_{\cdot 2i}$, and $\gamma_{\cdot i}$ represent the intercept, first slope, change in slope after the changepoint, and the changepoint location, respectively. Note, $x_{ji}$ is the time of measurement at measurement occasion $j$ for individual $i$. This model assumes that the measurement occasions are the same for both outcome variables. The model further allows for correlation between the random effects for the two related outcomes.  The distributional assumptions and priors for the BPREM are provided in Appendix~\ref{app:bprem_app}.

\subsection{BayesCREM}

The \fct{Bayes\_CREM} function fits a crossed random effects model (CREM) following \cite{Rohloff2024Identifiability}. The crossed random effects model (CREM) expands upon the REM by incorporating individual and group random effects. Therefore, the random coefficients are structured as follows,

\begin{equation} \label{eq:randcoeffcrem}
\beta_{qir} = \beta_q + b_{qi} + g_{qr}
\end{equation}

where $\beta_{qir}$ is the random coefficient for parameter $q$, $\beta_q$ is the fixed effect for parameter $q$, $b_{qi}$ is the individual random effect for parameter $q$, $g_{qr}$ is the group random effect for parameter $q$. The individual and group random effects are assumed to follow a multivariate normal distribution. The \pkg{BEND} package includes four different functional form options for the CREM: linear CREM (LCREM), quadratic CREM (QCREM), exponential CREM (ECREM), and piecewise CREM (PCREM). Each equation is presented below. In each equation, $Y_{jir}$ is the outcome, $x_{jir}$ is the time of measurement, and $\epsilon_{jir}$ is the error at measurement occasion $j$ for individual $i$ in group $r$. The distributional assumptions and priors for the CREMs are provided in Appendix~\ref{app:crem_app}.

\subsubsection{LCREM}
The LCREM model is
\begin{equation} \label{eq:lcrem}
Y_{jir} = \beta_{0ir} + \beta_{1ir}x_{jir} + \epsilon_{jir}
\end{equation}

where $\beta_{0ir}$ and $\beta_{1ir}$ are the intercept and slope, respectively. 

\subsubsection{QCREM}
The QCREM model is
\begin{equation} \label{eq:qcrem}
Y_{jir} = \beta_{0ir} + \beta_{1ir}x_{jir} + \beta_{2ir}x^2_{jir} + \epsilon_{jir}
\end{equation}

where $\beta_{0ir}$, $\beta_{1ir}$, and $\beta_{2ir}$ are the intercept, linear slope, and quadratic slope, respectively.

\subsubsection{ECREM}
The ECREM model is 
\begin{equation} \label{eq:ecrem}
Y_{jir} = \beta_{0ir} + \beta_{1ir}(1-e^{-\beta_{2ir}x_{jir}}) + \epsilon_{jir}
\end{equation}

where $\beta_{0ir}$, $\beta_{1ir}$, and $\beta_{2ir}$ are the intercept, total change, and growth rate, respectively. 

\subsubsection{PCREM}

\begin{equation} \label{eq:pcrem}
\begin{aligned} 
Y_{jir} = \beta_{0ir} + \beta_{1ir}x_{jir} + \beta_{2ir}(x_{jir}-\gamma_{ir})^+ + \epsilon_{jir} \\
(x_{jir}-\gamma_{ir})^+ = 
    \begin{cases}
      x_{jir}-\gamma_{ir} & \text{if $x_{jir}-\gamma_{ir}>0$}\\
      0 & \text{otherwise}\\
    \end{cases}  
\end{aligned}
\end{equation}

$\beta_{0ir}$, $\beta_{1ir}$, $\beta_{2ir}$, and $\gamma_{ir}$ are the intercept, first slope, change in slope after the changepoint, and changepoint location, respectively. 

%% -- Implementation and Examples ---------------------------------------------------------------

\section{Implementation and Examples} \label{sec:expl}

In this section, we will demonstrate how to implement the \pkg{BEND} functions in practice. Each example includes a simulated dataset, \proglang{R} code to apply the function, a summary of results, and plots of the observed and fitted data. To make these examples more accessible to readers, the datasets used here are available in the \pkg{BEND} package, so readers can follow along and run these examples on their own. To begin, install and load the \pkg{BEND} package using the following code:

\begin{CodeChunk}
\begin{CodeInput}
R> install.packages("BEND")
R> library(BEND)
R> set.seed(1)
\end{CodeInput}
\end{CodeChunk}

\pkg{BEND} will automatically install the packages \pkg{coda} \citep{coda}, \pkg{label.switching} \citep{label}, and \pkg{rjags} \citep{rjags} as dependencies. \pkg{rjags} requires the installation of \proglang{JAGS} software which is freely available online: \hyperlink{https://mcmc-jags.sourceforge.io/}{https://mcmc-jags.sourceforge.io/}. 

For simplicity, each example below uses a simulated dataset. For examples of how the \pkg{BEND} functions can be applied to real-world datasets, please see the original literature for each function: for PREMM see \cite{Lock2018Detecting}, for CI-PREMM see \cite{Lamm2022Incorporation}, for BPREM see \cite{Peralta2022Bayesian}, for CREM see \cite{Rohloff2024Identifiability}. 

\subsection{BayesPREM}

First, load and preview the observed data:

\begin{CodeChunk}
\begin{CodeInput}
R> data(SimData_PREM)
R> head(SimData_PREM)

id time     y class_pred_1 class_pred_2 outcome_pred_1
 1    0 10.56            0            1           7.97
 1    1  9.84            0            1           5.08
 1    2 10.66            0            1           7.59
 1    3 11.06            0            1           7.12
 1    4 11.03            0            1           5.87
 1    5 11.54            0            1           9.92
\end{CodeInput}
\end{CodeChunk}

\begin{figure}[h]
\centering
\includegraphics{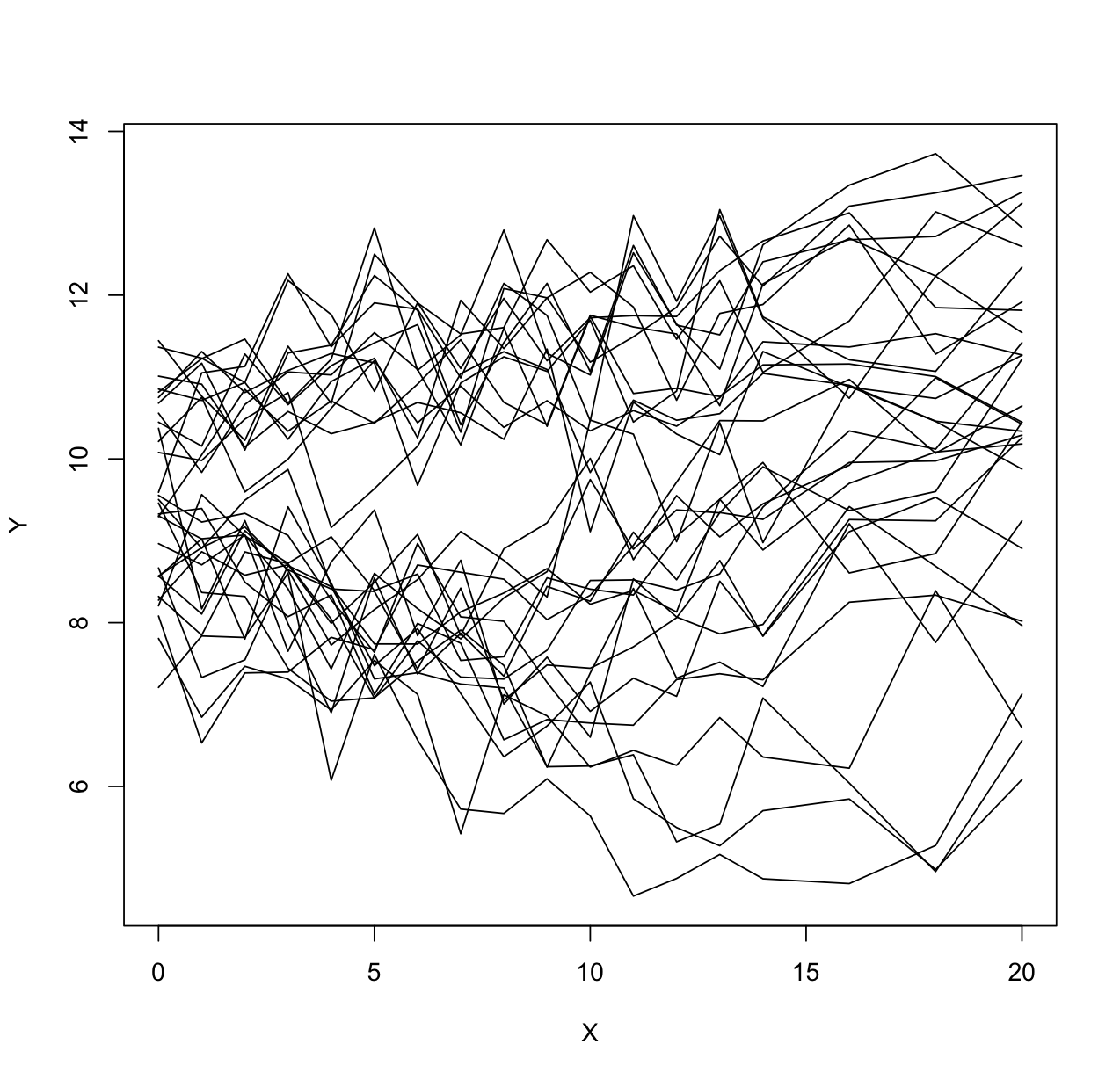}
\caption{\label{fig:prem_obs} Simulated PREM Data}
\end{figure}

Note, the \pkg{BEND} package assumes data follows a long format where each row is a different measurement occasion. To plot the observed data, use {\tt plot\_BEND()} as shown below (output presented in Figure~\ref{fig:prem_obs}). This function allows users to generate a plot of observed trajectories following the same format and style as the fitted plots obtained by applying the generic \fct{plot} function to fitted \pkg{BEND} model objects (e.g., see Figure~\ref{fig:prem_fit}). 

\begin{CodeChunk}
\begin{CodeInput}
R> plot_BEND(data = SimData_PREM,
+            id_var = "id",
+            time_var = "time",
+            y_var = "y")
\end{CodeInput}
\end{CodeChunk}

For now, we will only show examples of how to fit the PREM and CI-PREMM using the \fct{Bayes\_PREM} function.

\subsubsection{PREM}

The PREM is the default model in \fct{Bayes\_PREM}. Therefore, to fit the PREM we only need to specify the dataset and provide the variable names for the individual identifiers (\code{id_var}), time (\code{time_var}), and outcome (\code{y_var}). By default, the maximum number of changepoints considered by the model is set to 2, but this can be changed with the argument \code{max_cp} (e.g., \code{max_cp} = 4). The following is the corresponding code and output for \fct{Bayes\_PREM}.

\begin{CodeChunk}
\begin{CodeInput}
R> results_prem <- Bayes_PREM(data = SimData_PREM, 
+                             id_var = "id", 
+                             time_var = "time", 
+                             y_var = "y")

R> summary(results_prem)

Class Dependent Parameters:
                                Estimate Lower CI Upper CI
Class 1: Intercept Mean            9.727    9.282   10.176
Class 1: Slope Mean               -0.233   -0.278   -0.187
Class 1: Changepoint 1 Mean        2.823    1.135    5.033
Class 1: Change in Slope 1 Mean    0.352    0.301    0.400
Class 1: Intercept Var             1.439    0.839    2.347
Class 1: Slope Var                 0.001    0.000    0.004
Class 1: Changepoint 1 Var        20.845   13.921   24.840
Class 1: Change in Slope 1 Var     0.003    0.000    0.007

Class Independent Parameters:
          Estimate Lower CI Upper CI
Error Var    0.345    0.303    0.391

Gelman's msrf: 1.09 
Mean psrf: 1.01 
DIC: 1033
\end{CodeInput}
\end{CodeChunk}

The \fct{summary} function provides the estimates and corresponding 95 percent credible intervals for each parameter. It also includes information about model convergence (i.e., Gelman's MSRF, Mean PSRF) and fit (i.e., DIC). By default, \fct{summary} will print the parameter estimates for the number of changepoints with the largest probability. The probability distribution for the different latent number of changepoints $\mathcal{K}$ can be obtained using the \fct{getKProb} function:

\begin{CodeChunk}
\begin{CodeInput}
R> getKProb(results_prem)

K (Number of Changepoints) Probabilities

Class 1
    Probability
K=0       0.000
K=1       0.986
K=2       0.014
\end{CodeInput}
\end{CodeChunk}

For additional details on other extraction methods available in BEND, see Appendix~\ref{app:extract_methods}.

\begin{figure}[h]
\centering
\includegraphics{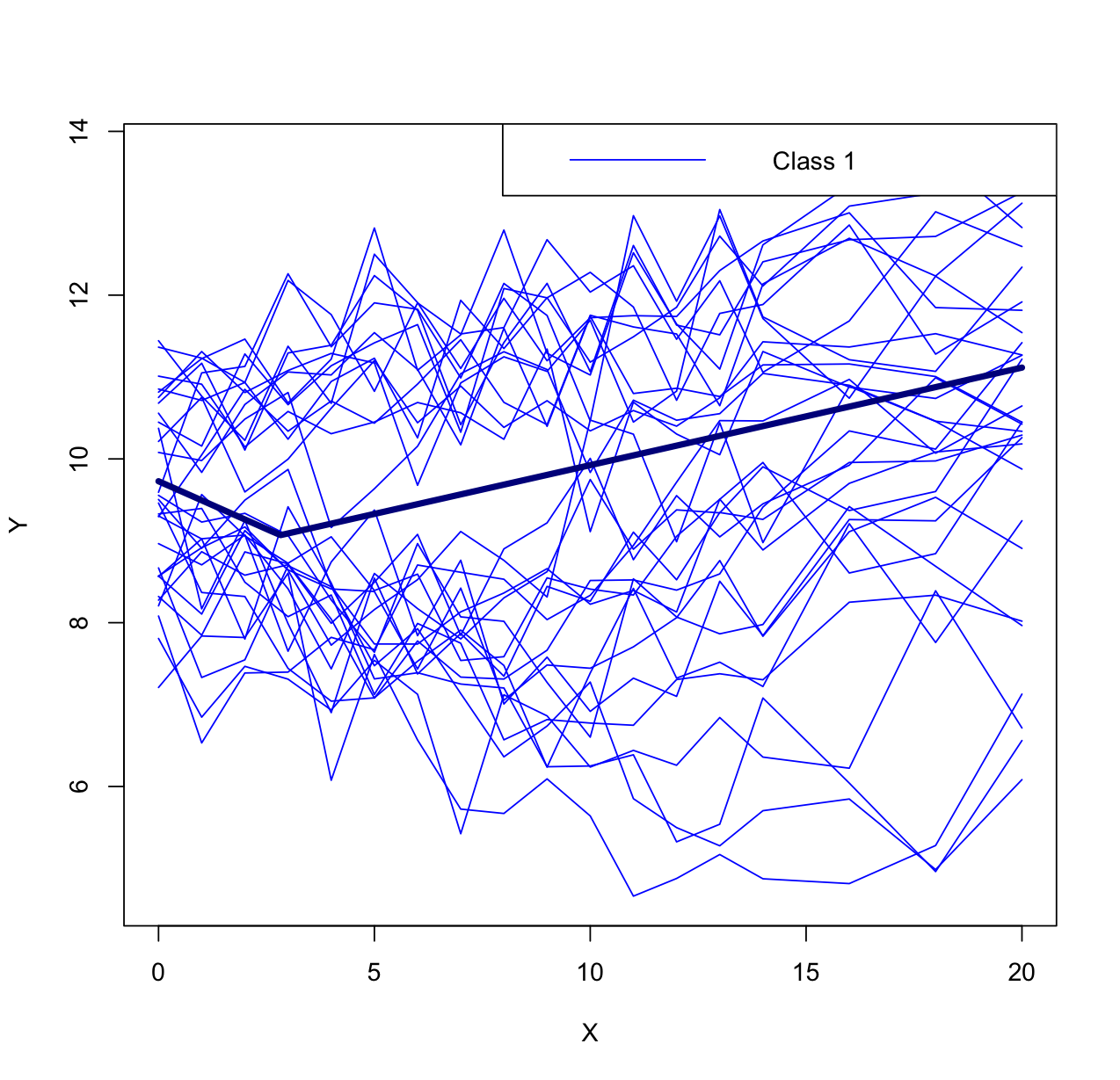}
\caption{\label{fig:prem_fit} Fitted Results for the PREM }
\end{figure}

To plot the fitted results, users can utilize the \fct{plot} function and supply the fitted \fct{Bayes\_PREM} model results (output shown in Figure~\ref{fig:prem_fit}).

\begin{CodeChunk}
\begin{CodeInput}
R> plot(results_prem)
\end{CodeInput}
\end{CodeChunk}

The results of fitting the PREM in \fct{Bayes\_PREM} suggest that there is only one changepoint ($P(\mathcal{K}=1)=0.986$). Individuals, on average, will have a negative growth trajectory until $x=2.823$ ($\gamma_1$) at which point the growth trajectory increases. There is a great deal of variation in the place this changepoint occurs ($\sigma_{b_{\gamma_1}}^2=20.845$).

In the interest of brevity, full examples for the CI-PREM and PREMM are not provided. However, both can be easily specified by making minor modifications to the \fct{Bayes\_PREM} function arguments. To fit the CI-PREM, we need to add the variable names for the outcome predictive covariates (\code{outcome_predictive_vars}). This argument can take either a single value or a vector of variable names:

\begin{CodeChunk}
\begin{CodeInput}
R> results_ciprem <- Bayes_PREM(data = SimData_PREM,
+                               id_var = "id",
+                               time_var = "time",
+                               y_var = "y",
+                               outcome_predictive_vars = "outcome_pred_1")
\end{CodeInput}
\end{CodeChunk}

If it is hypothesized that there are latent classes with different patterns of growth, one can fit the PREMM in by setting \code{n_class} equal to the number of hypothesized latent classes (can be any value greater than 1):

\begin{CodeChunk}
\begin{CodeInput}
R> results_premm <- Bayes_PREM(data = SimData_PREM,
+                              id_var = "id",
+                              time_var = "time",
+                              y_var = "y",
+                              n_class = 2)
\end{CodeInput}
\end{CodeChunk}

The \fct{Bayes\_PREM} function also allows for different options for the set of priors for $\mathcal{K}$ and the scaling parameter on the random coefficients. See Appendices~\ref{app:prem_app},~\ref{app:ciprem_app},~\ref{app:premm_app}, and~\ref{app:cipremm_app} for more information.

\subsubsection{CI-PREMM}

To determine which characteristics may impact class membership and predict the outcome, we can fit the CI-PREMM in \fct{Bayes\_PREM}. To do this, additional arguments need to be specified. In addition to the individual identifier, time, and outcome variables, the user must set \code{n_class=2} (this must be equal to 2 for the CI-PREMM) and provide the names of the class and outcome predictive variables. 

\begin{CodeChunk}
\begin{CodeInput}
R> results_cipremm <- Bayes_PREM(data = SimData_PREM,
+                              id_var = "id",
+                              time_var = "time",
+                              y_var = "y",
+                              n_class = 2,
+                              class_predictive_vars = c("class_pred_1",       
+                                                        "class_pred_2"),
+                              outcome_predictive_vars = "outcome_pred_1")

R> summary(results_cipremm)

Class Dependent Parameters:
                                Estimate Lower CI Upper CI
Class 1: Intercept Mean            8.502    8.077    8.918
Class 1: Slope Mean               -0.215   -0.257   -0.170
Class 1: Changepoint 1 Mean        7.961    5.476   10.212
Class 1: Change in Slope 1 Mean    0.386    0.331    0.437
Class 1: Intercept Var             0.335    0.122    0.836
Class 1: Slope Var                 0.001    0.000    0.006
Class 1: Changepoint 1 Var        14.497    5.881   24.351
Class 1: Change in Slope 1 Var     0.001    0.000    0.009
Class 2: Intercept Mean           10.180    9.520   10.619
Class 2: Slope Mean                0.076    0.045    0.108
Class 2: Changepoint 1 Mean                               
Class 2: Change in Slope 1 Mean                           
Class 2: Intercept Var             0.333    0.030    1.746
Class 2: Slope Var                 0.002    0.001    0.007
Class 2: Changepoint 1 Var                                
Class 2: Change in Slope 1 Var                            

Class Independent Parameters:
                                 Estimate Lower CI Upper CI
Error Var                           0.331    0.291    0.377
outcome_pred_1                      0.053    0.014    0.096
class_pred_1 (in log-odds units)   -4.148   -8.218   -1.243
class_pred_2 (in log-odds units)   -0.692   -2.458    1.013
Logistic Intercept                  0.623   -0.454    1.805

Gelman's msrf: 2.76 
Mean psrf: 1.13 
DIC: 1005
\end{CodeInput}
\end{CodeChunk}

Again, to view the probability distribution for the different number of changepoints we can look at the model output. Each class will have its own probability distribution for the number of changepoints:

\begin{CodeChunk}
\begin{CodeInput}
R> getKProb(results_cipremm)

K (Number of Changepoints) Probabilities

Class 1
    Probability
K=0       0.000
K=1       0.975
K=2       0.025

Class 2
    Probability
K=0       0.980
K=1       0.020
K=2       0.000
\end{CodeInput}
\end{CodeChunk}

To get the proportion of individuals in each class:

\begin{CodeChunk}
\begin{CodeInput}
R> class_probs <- getClassProb(results_cipremm)
R> prop.table(table(class_probs[["individ_class_info"]][["class_membership"]]))

    1     2 
0.567 0.433
\end{CodeInput}
\end{CodeChunk}

\begin{figure}[h]
\centering
\includegraphics{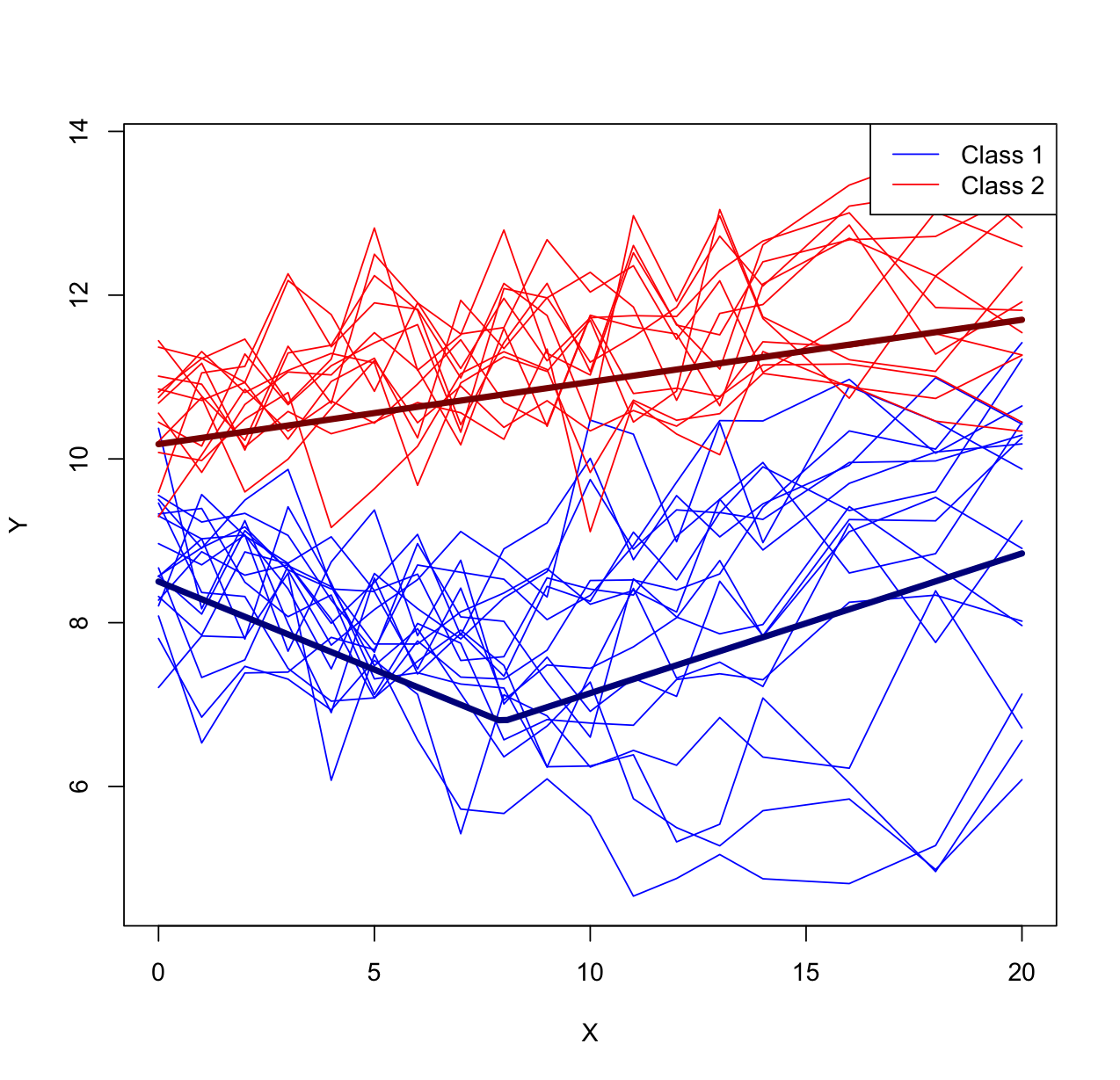}
\caption{\label{fig:cipremm_fit} Fitted Results for the CI-PREMM}
\end{figure}

Again, to plot the fitted results use {\tt plot()} (output shown in Figure~\ref{fig:cipremm_fit}).

\begin{CodeChunk}
\begin{CodeInput}
R> plot(results_cipremm)
\end{CodeInput}
\end{CodeChunk}

The empirical class probabilities represent the percent of individuals that belong to each class. In this example, slightly more individuals belong to class 1 ($56.7\%$). These results also suggest that the number of changepoints and therefore, the shape of the growth trajectory, differs by class. Class 1 follows a piecewise trajectory with one changepoint where there is an initial decrease in growth until $x=7.961$, then the growth trajectory increases. Class 2, on the other hand, does not have any changepoints, meaning it follows a constant (linear) growth trajectory.  

The class-predictive covariates allow us to gain insight on the characteristics that impact class membership. In this model, we are predicting the log-odds of being assigned to Class 2 (Class 1 is the reference group). Both class-predictive covariates are negative indicating that belonging to the focal group of these variables decreases the odds of being assigned to Class 2.  For example, in this simulated example, \code{class_pred_2} is a (0/1) variable denoting membership to a certain group. Based on the parameter estimate for \code{class_pred_2}, individuals that belong to this group are about 2 times less likely to be assigned to Class 2 than those who do not belong to this group. Lastly, the addition of the outcome predictive covariate in the model explains some of the variation in the outcome (the error variance has decreased from the PREM results). For more detailed information about how to interpret parameters from the \fct{Bayes\_PREM} models see the original papers, \cite{Lock2018Detecting} and \cite{Lamm2022Incorporation}. 

\subsection{BayesBPREM}

Fitting the BPREM in \pkg{BEND} will follow similar steps to \fct{Bayes\_PREM}. First, load the observed data:

\begin{CodeChunk}
\begin{CodeInput}
R> data(SimData_BPREM)
R> head(SimData_BPREM)

id  time      y1      y2
 1     0   33.01   40.74
 1     1  124.83   77.58
 1     2  153.22  108.86
 1     3  167.50  131.91
 1     4  149.69  134.36
 1     5  157.28  159.20
\end{CodeInput}
\end{CodeChunk}

\begin{figure}[h]
\centering
\includegraphics{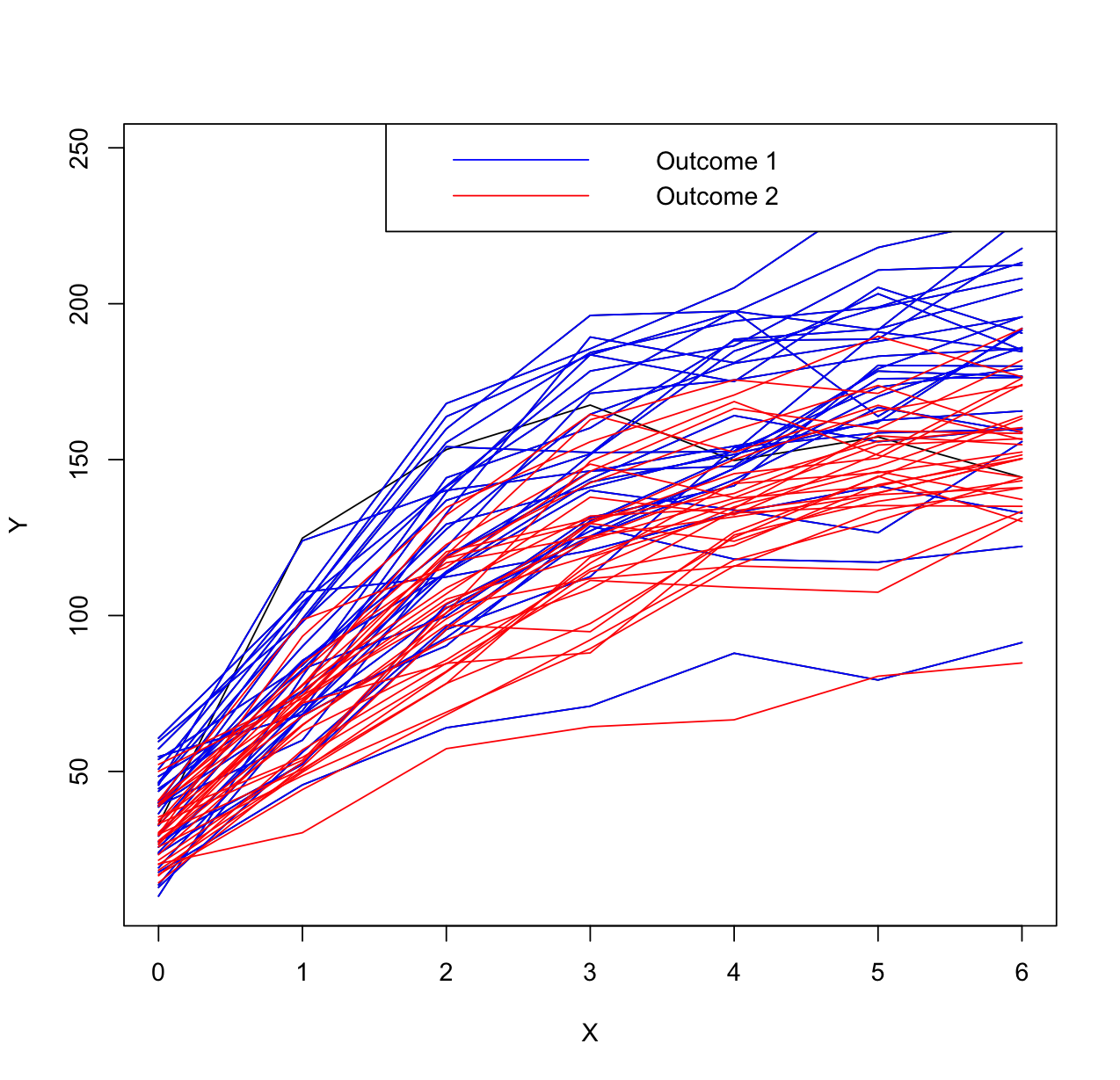}
\caption{\label{fig:bprem_obs} Simulated BPREM Data}
\end{figure}

Plot the observed data using {\tt plot\_BEND()} (output shown in Figure~\ref{fig:bprem_obs}).

\begin{CodeChunk}
\begin{CodeInput}
R> plot_BEND(data = SimData_BPREM,
+            id_var = "id",
+            time_var = "time",
+            y_var = "y1",
+            y2_var = "y2")
\end{CodeInput}
\end{CodeChunk}

To fit the BPREM, we only need to specify the dataset and provide the variable names for the individual identifiers (\code{id_var}), time (\code{time_var}), and the two outcomes (\code{y1_var} and \code{y2_var}).

The \fct{summary} function provides the parameter estimates and credible intervals of the growth parameters for both outcomes, as well as the covariance and correlation matrices for the random effects and error variances (continued below). To interpret the relationship between the two outcomes, we recommend focusing on the correlation matrix as that is a standardized measure of interdependence. The \fct{summary} function also reports model convergence and fit. 

\begin{CodeChunk}
\begin{CodeInput}
R> results_bprem <- Bayes_BPREM(data = SimData_BPREM,
+                               id_var = "id",
+                               time_var = "time",
+                               y1_var = "y1",
+                               y2_var = "y2")

R> summary(results_bprem)

Bayesian bivariate piecewise random effects model
Data: SimData_BPREM 
Outcomes: y1 y2 

Fixed Effect Parameters:
                                Estimate Lower CI Upper CI
Outcome 1: Intercept Mean          37.68    33.31    42.27
Outcome 1: Slope Mean              44.85    40.59    49.76
Outcome 1: Change in Slope Mean   -35.04   -39.79   -30.48
Outcome 1: Changepoint Mean         2.64     2.26     3.00
Outcome 2: Intercept Mean          30.98    27.45    34.64
Outcome 2: Slope Mean              34.15    30.47    37.60
Outcome 2: Change in Slope Mean   -25.07   -29.34   -20.66
Outcome 2: Changepoint Mean         2.77     2.42     3.23

Random Effect Parameters:
Variances:
                               Estimate Lower CI Upper CI
Outcome 1: Intercept Var         95.229   27.667   214.60
Outcome 1: Slope Var            129.757   41.627   268.89
Outcome 1: Change in Slope Var  128.260   39.856   271.10
Outcome 1: Changepoint Var        0.565    0.184     1.20
Outcome 2: Intercept Var         56.900   17.753   126.33
Outcome 2: Slope Var             55.188   20.723   109.96
Outcome 2: Change in Slope Var   80.610   32.554   160.11
Outcome 2: Changepoint Var        0.312    0.001     1.16

Covariances:
                                                            Estimate Lower CI Upper CI
Cov(Outcome 1: Slope, Outcome 1: Intercept)                   73.849   22.535  146.085
Cov(Outcome 1: Change in Slope, Outcome 1: Intercept)        -70.755 -141.547  -22.311
Cov(Outcome 1: Change in Slope, Outcome 1: Slope)           -107.527 -239.534  -24.384
Cov(Outcome 1: Changepoint, Outcome 1: Intercept)             -4.598   -9.778   -0.816
Cov(Outcome 1: Changepoint, Outcome 1: Slope)                 -6.742  -15.381   -0.877
Cov(Outcome 1: Changepoint, Outcome 1: Change in Slope)        5.837    0.175   13.761
Cov(Outcome 2: Intercept, Outcome 1: Intercept)               51.721   16.182  112.511
Cov(Outcome 2: Intercept, Outcome 1: Slope)                   60.786   15.232  129.125
Cov(Outcome 2: Intercept, Outcome 1: Change in Slope)        -54.226 -120.408  -11.575
Cov(Outcome 2: Intercept, Outcome 1: Changepoint)             -3.848   -8.469   -0.523
Cov(Outcome 2: Slope, Outcome 1: Intercept)                   28.741   -8.460   75.035
Cov(Outcome 2: Slope, Outcome 1: Slope)                       40.352    1.883   96.236
Cov(Outcome 2: Slope, Outcome 1: Change in Slope)            -21.011  -73.846   19.388
Cov(Outcome 2: Slope, Outcome 1: Changepoint)                 -2.461   -6.088    0.200
Cov(Outcome 2: Slope, Outcome 2: Intercept)                   27.589   -2.039   60.385
Cov(Outcome 2: Change in Slope, Outcome 1: Intercept)        -34.604  -90.192    6.963
Cov(Outcome 2: Change in Slope, Outcome 1: Slope)            -43.358 -107.225    1.422
Cov(Outcome 2: Change in Slope, Outcome 1: Change in Slope)   22.486  -25.081   83.743
Cov(Outcome 2: Change in Slope, Outcome 1: Changepoint)        3.085    0.087    7.331
Cov(Outcome 2: Change in Slope, Outcome 2: Intercept)        -35.932  -77.208   -3.888
Cov(Outcome 2: Change in Slope, Outcome 2: Slope)            -55.872 -114.338  -19.643
Cov(Outcome 2: Changepoint, Outcome 1: Intercept)             -1.193   -5.897    1.933
Cov(Outcome 2: Changepoint, Outcome 1: Slope)                 -1.555   -6.484    1.631
Cov(Outcome 2: Changepoint, Outcome 1: Change in Slope)        1.594   -1.470    6.465
Cov(Outcome 2: Changepoint, Outcome 1: NA)                     0.126   -0.091    0.478
Cov(Outcome 2: Changepoint, Outcome 2: Intercept)             -1.161   -4.819    1.152
Cov(Outcome 2: Changepoint, Outcome 2: Slope)                 -2.276   -8.687    0.586
Cov(Outcome 2: Changepoint, Outcome 2: Change in Slope)        1.503   -1.507    6.460

Correlations:
                                                             Estimate Lower CI Upper CI
Corr(Outcome 1: Slope, Outcome 1: Intercept)                    0.714    0.234    0.957
Corr(Outcome 1: Change in Slope, Outcome 1: Intercept)         -0.689   -0.938   -0.240
Corr(Outcome 1: Change in Slope, Outcome 1: Slope)             -0.812   -0.946   -0.493
Corr(Outcome 1: Changepoint, Outcome 1: Intercept)             -0.663   -0.935   -0.138
Corr(Outcome 1: Changepoint, Outcome 1: Slope)                 -0.755   -0.947   -0.248
Corr(Outcome 1: Changepoint, Outcome 1: Change in Slope)        0.661    0.036    0.929
Corr(Outcome 2: Intercept, Outcome 1: Intercept)                0.730    0.342    0.947
Corr(Outcome 2: Intercept, Outcome 1: Slope)                    0.727    0.280    0.946
Corr(Outcome 2: Intercept, Outcome 1: Change in Slope)         -0.657   -0.950   -0.185
Corr(Outcome 2: Intercept, Outcome 1: Changepoint)             -0.696   -0.951   -0.138
Corr(Outcome 2: Slope, Outcome 1: Intercept)                    0.418   -0.123    0.797
Corr(Outcome 2: Slope, Outcome 1: Slope)                        0.487    0.029    0.817
Corr(Outcome 2: Slope, Outcome 1: Change in Slope)             -0.250   -0.673    0.250
Corr(Outcome 2: Slope, Outcome 1: Changepoint)                 -0.455   -0.806    0.039
Corr(Outcome 2: Slope, Outcome 2: Intercept)                    0.536   -0.036    0.876
Corr(Outcome 2: Change in Slope, Outcome 1: Intercept)         -0.413   -0.775    0.091
Corr(Outcome 2: Change in Slope, Outcome 1: Slope)             -0.430   -0.769    0.017
Corr(Outcome 2: Change in Slope, Outcome 1: Change in Slope)    0.221   -0.268    0.644
Corr(Outcome 2: Change in Slope, Outcome 1: Changepoint)        0.475    0.012    0.817
Corr(Outcome 2: Change in Slope, Outcome 2: Intercept)         -0.566   -0.877   -0.057
Corr(Outcome 2: Change in Slope, Outcome 2: Slope)             -0.836   -0.956   -0.594
Corr(Outcome 2: Changepoint, Outcome 1: Intercept)             -0.342   -0.961    0.432
Corr(Outcome 2: Changepoint, Outcome 1: Slope)                 -0.392   -0.983    0.396
Corr(Outcome 2: Changepoint, Outcome 1: Change in Slope)        0.399   -0.347    0.961
Corr(Outcome 2: Changepoint, Outcome 1: NA)                     0.430   -0.376    0.955
Corr(Outcome 2: Changepoint, Outcome 2: Intercept)             -0.395   -0.944    0.414
Corr(Outcome 2: Changepoint, Outcome 2: Slope)                 -0.485   -0.923    0.463
Corr(Outcome 2: Changepoint, Outcome 2: Change in Slope)        0.323   -0.548    0.887

Error:
Variance-Covariance:
                                        Estimate Lower CI Upper CI
Outcome 1: Error Var                        91.9    70.04    118.5
Outcome 2: Error Var                        57.6    40.64     76.9
Cov(Outcome 1: Error, Outcome 2: Error)     11.7    -1.88     26.5

Correlation:
                                         Estimate Lower CI Upper CI
Corr(Outcome 1: Error, Outcome 2: Error)    0.161   -0.026    0.346

Gelman's msrf: 1.97 
Mean psrf: 1.11 
DIC: 3139
\end{CodeInput}
\end{CodeChunk}

\begin{figure}[h]
\centering
\includegraphics{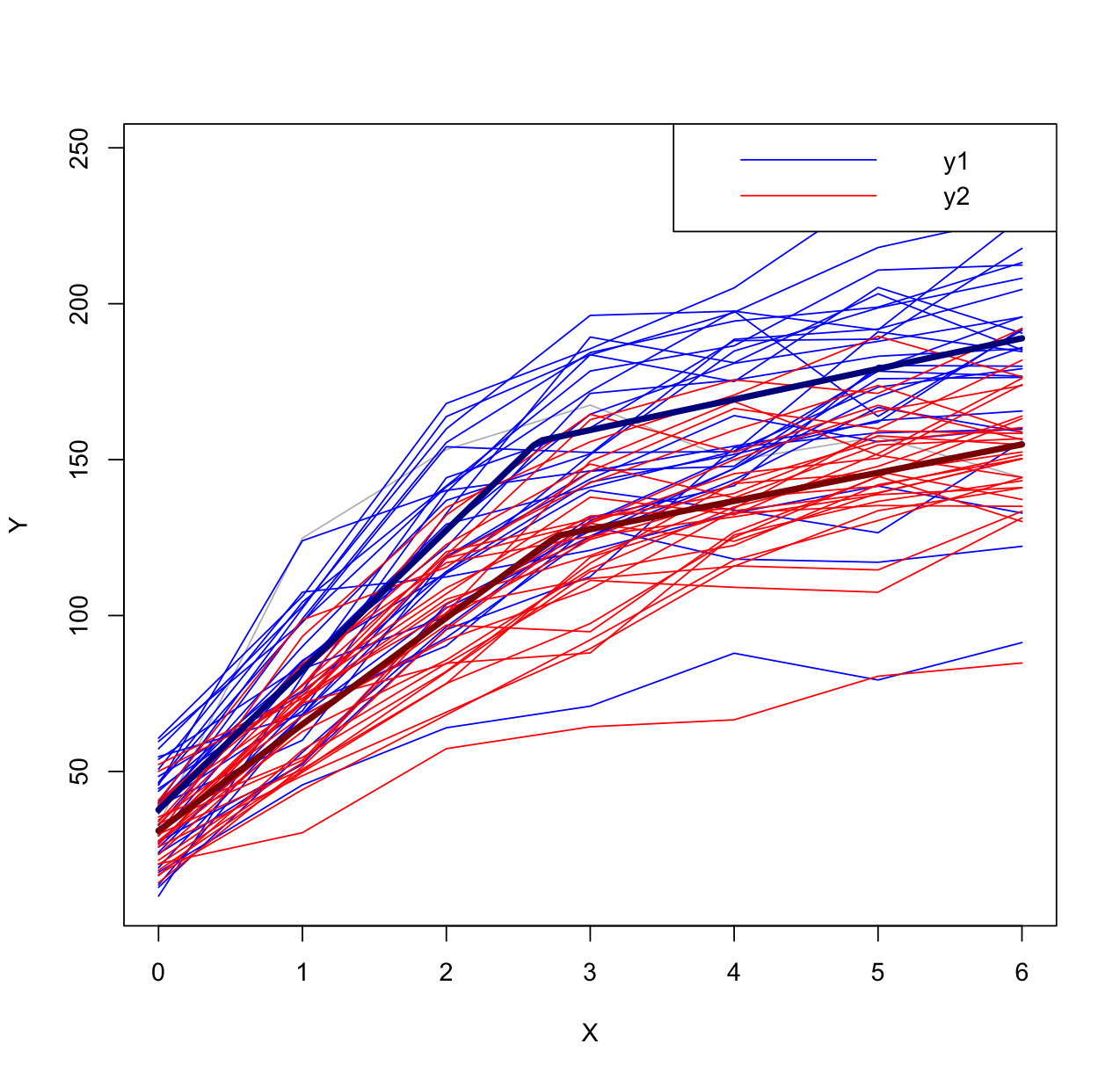}
\caption{\label{fig:bprem_fit} Fitted Results for the BPREM}
\end{figure}

To plot the fitted results, again use {\tt plot())} (output shown in Figure~\ref{fig:bprem_fit}).

\begin{CodeChunk}
\begin{CodeInput}
R> plot(results_bprem)
\end{CodeInput}
\end{CodeChunk}

According to these results, both outcomes follow a similar growth trajectory with a faster rate of growth before the changepoint, followed by a slower rate of growth. The correlation of the error terms suggests a moderate relationship between the two outcomes and the random effect correlation matrix suggests there is some interdependence among the two outcomes’ growth trajectories at different stages of development. For example, the two outcomes were strongly correlated at the first timepoint ($\hat{\rho}_{b_{20}b_{10}}=0.730$) and moderately correlated during their first phase of growth ($\hat{\rho}_{b_{21}b_{11}}=0.487$). But the strength of the relationship decreased after the changepoint ($\hat{\rho}_{b_{22}b_{12}}=0.221$). This section provides only a brief summary of the results. Readers are encouraged to review the original paper to get an in-depth example of how to interpret the BPREM parameters \citep[see][]{Peralta2022Bayesian}.

\subsection{BayesCREM}

Lastly, we will review how to fit a PCREM using \fct{Bayes\_CREM}. First, load the observed data:

\begin{CodeChunk}
\begin{CodeInput}
R> data(SimData_PCREM)
R> head(SimData_PCREM)

id  teacherid  time       y
 1         43   0.0   35.41
 1         10   0.5   46.25
 1         31   1.0   74.84
 1         35   1.5   82.17
 1         39   3.5  153.11
 1         14   5.5  163.09
\end{CodeInput}
\end{CodeChunk}

\begin{figure}[h]
\centering
\includegraphics{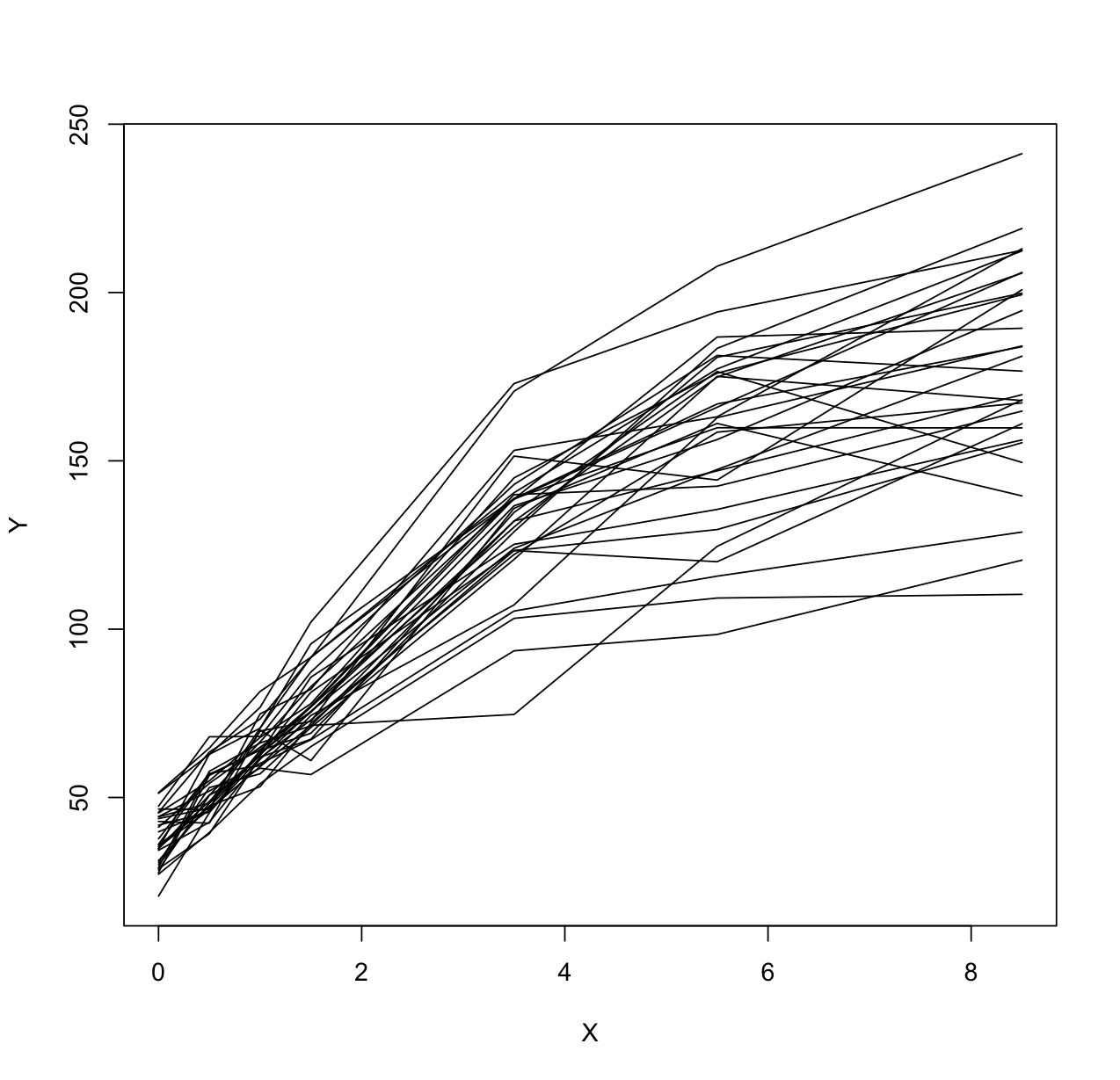}
\caption{\label{fig:pcrem_obs} Simulated PCREM Data}
\end{figure}

Plot the observed data, shown in Figure~\ref{fig:pcrem_obs}.

\begin{CodeChunk}
\begin{CodeInput}
R> plot_BEND(data = SimData_PCREM,
+            id_var = "id",
+            time_var = "time",
+            y_var = "y")
\end{CodeInput}
\end{CodeChunk}

To fit a PCREM, we need to specify the dataset, provide the variable names for the individual identifiers (\code{id_var}), the identifiers for the crossed factor (\code{cross_id_var}), time (\code{time_var}), and outcome (\code{y_var}), and indicate the desired functional form (\code{form=“piecewise”}). The four acceptable arguments for \code{form} are \code{“linear”},  \code{“quadratic”}, \code{“exponential”}, and \newline \code{“piecewise”}.

\begin{CodeChunk}
\begin{CodeInput}
R> results_pcrem <- Bayes_CREM(data = SimData_PCREM,
+                              ind_id_var = "id",
+                              cross_id_var = "teacherid",
+                              time_var = "time",
+                              y_var = "y",
+                              form="piecewise")

R> summary(results_pcrem)

Bayesian crossed random effects model
Data: SimData_PCREM 
Outcome: y 
Individuals: id 
Group: teacherid 

Fixed Effect Parameters:
                     Estimate Lower CI Upper CI
Intercept Mean          37.11    34.51    39.63
Slope Mean              28.70    26.78    30.60
Change in Slope Mean   -21.99   -24.54   -19.55
Changepoint Mean         3.73     3.19     4.26

Random Effect Parameters:
Individual Random Effects Variance-Covariance:
                                  Estimate Lower CI Upper CI
Intercept Var                       23.184    9.434   45.723
Slope Var                           13.110    3.818   28.035
Change in Slope Var                 10.928    0.101   34.645
Changepoint Var                      0.794    0.187    1.840
Cov(Intercept, Slope)                4.655   -4.122   13.219
Cov(Intercept, Change in Slope)      1.050   -7.851   13.429
Cov(Slope, Change in Slope)         -4.454  -17.073    2.323
Cov(Intercept, Changepoint)         -0.582   -2.982    1.700
Cov(Slope, Changepoint)              0.026   -2.389    1.925
Cov(Change in Slope, Changepoint)   -1.339   -4.469    0.379

Group Random Effects Variance-Covariance:
                                  Estimate Lower CI Upper CI
Intercept Var                       18.256    6.185   36.774
Slope Var                            4.781    0.188   13.202
Change in Slope Var                  6.950    0.014   29.247
Changepoint Var                      0.343    0.002    1.141
Cov(Intercept, Slope)               -2.521  -11.096    2.230
Cov(Intercept, Change in Slope)      0.195   -9.161   10.554
Cov(Slope, Change in Slope)         -0.660   -7.917    3.430
Cov(Intercept, Changepoint)          0.988   -0.345    3.443
Cov(Slope, Changepoint)             -0.492   -2.363    0.298
Cov(Change in Slope, Changepoint)   -0.286   -2.291    0.867

Error Variance:
          Estimate Lower CI Upper CI
Error Var     13.4     8.09     20.9

Gelman's msrf: 1.26 
Mean psrf: 1.03 
DIC: 1287
\end{CodeInput}
\end{CodeChunk}

\begin{figure}[h]
\centering
\includegraphics{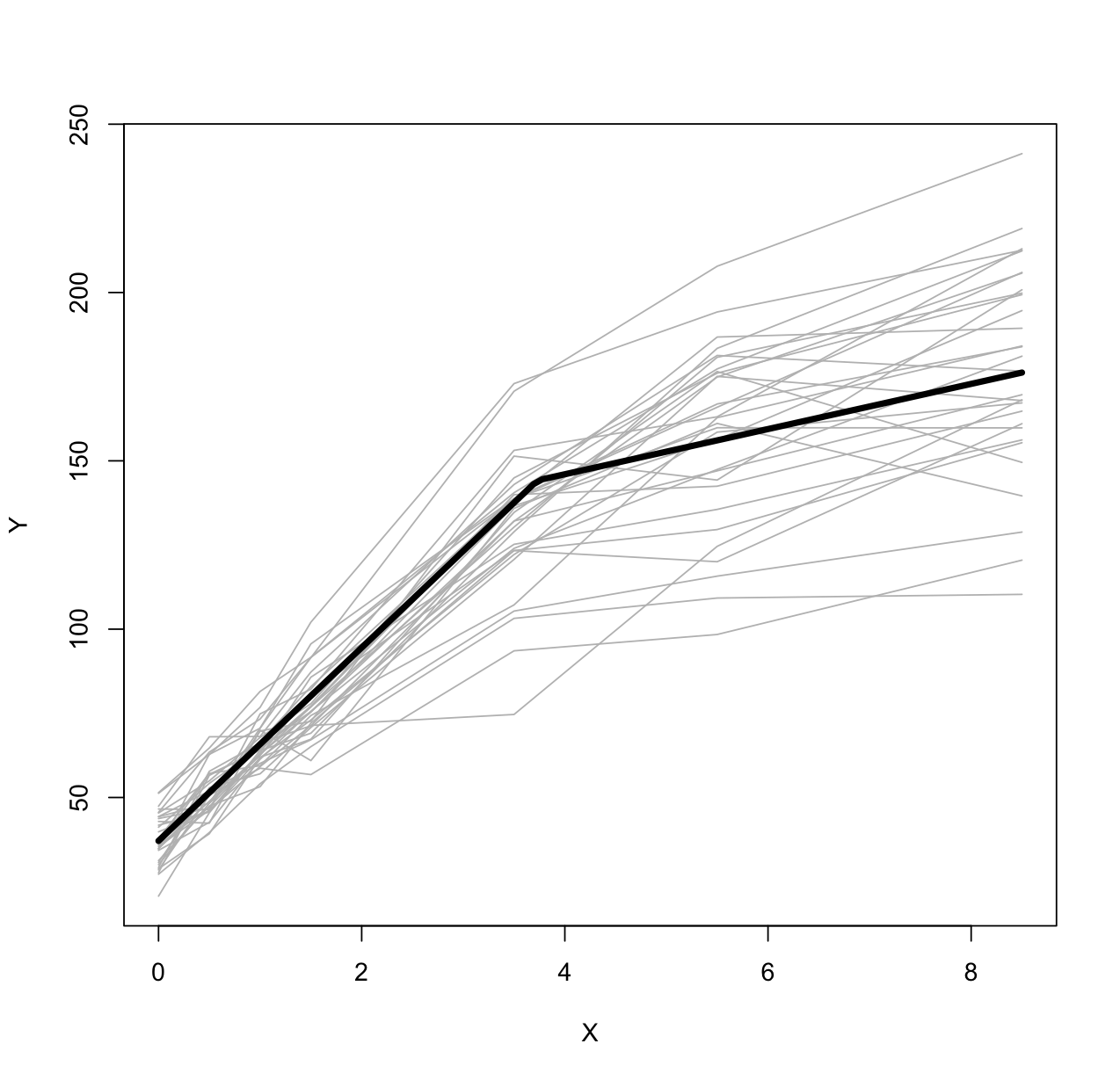}
\caption{\label{fig:pcrem_fit} Fitted Results for the PCREM}
\end{figure}

The \fct{summary} function provides the parameter estimates and credible intervals for the growth parameters, as well as the covariance matrices for the individual and group random effects. To plot the fitted results, use {\tt plot()} (output shown in Figure~\ref{fig:pcrem_fit}).

\begin{CodeChunk}
\begin{CodeInput}
R> plot(results_pcrem)
\end{CodeInput}
\end{CodeChunk}

These results suggest that there is an initial rate of growth which slows after $x=3.73$. Based on the random effects covariance matrices, there appears to be more variation in the growth parameters due to the individuals than groups. Again, please see the original paper for more guidance on how to interpret the PCREM parameters \citep[see][]{Rohloff2024Identifiability}. 

% -- Discussion ---------------------------------------------------------------

\section{Discussion} \label{sec:discuss}

There are a few limitations and notes to address about the \pkg{BEND} functions. First, the functions differ in whether they estimate the covariances between the random effects parameters. Both \fct{Bayes\_BPREM} and \fct{Bayes\_CREM} report covariances, but \fct{Bayes\_PREM} does not. This is something that can be improved upon in future versions of the \proglang{R} package. 

Second, we recommend using a mean PSRF of 1.2 as a cutoff for convergence \citep{Lock2018Detecting, Peralta2022Bayesian, Rohloff2024Identifiability}, but a smaller value might be desired. To help with convergence, the user can adjust the number of iterations for burn-in and posterior sampling from their defaults by using the following function arguments: \code{iters\_adapt}, \code{iters\_burn\_in}, \code{iters\_sampling}, and \code{thin}. Each \pkg{BEND} function has a different default for these arguments. See the \pkg{BEND} package documentation for this information. 

Lastly, the functions assume that each individual in the dataset has the same number of measurement occasions. In other words, in the long data form, every individual should have the same number of rows. The outcome value can be missing if that measurement occasion was not collected, but there must still be a row for that missing measurement occasion. This is another detail that can be improved upon in future versions of the package. 

The goal of this paper was to introduce and familiarize readers with the \proglang{R} package-Bayesian Estimation of Nonlinear Data (\pkg{BEND}). We presented three functions from the package that can be used for modeling nonlinear longitudinal data. \fct{Bayes\_PREM} allows researchers to empirically identify the number and location of random changepoints in a piecewise random effects model. Extensions of this function provide options for modeling multiple latent classes with different longitudinal growth patterns and incorporating covariates to predict the outcome and latent class membership. \fct{Bayes\_BPREM} allows researchers to jointly model the longitudinal piecewise trajectories of two interrelated outcomes. Lastly, \fct{Bayes\_CREM} allows researchers to estimate the impact of group membership on longitudinal growth. Both linear and nonlinear functional forms are available. 

\pkg{BEND} provides a user-friendly method of estimating complex longitudinal models. By making these functions available in \proglang{R} Software, we can ensure that these methods are accessible by anyone wishing to answer these types of research questions. All the code is open source, so users are free to see how our methods are applied and modify as necessary for their unique purposes. 

%% -- Bibliography -------------------------------------------------------------
%% - References need to be provided in a .bib BibTeX database.
%% - All references should be made with \cite, \citet, \citep, \citealp etc.
%%   (and never hard-coded). See the FAQ for details.
%% - JSS-specific markup (\proglang, \pkg, \code) should be used in the .bib.
%% - Titles in the .bib should be in title case.
%% - DOIs should be included where available.

\newpage

\bibliography{refs}

%% -- Appendices --------------------------------------------------------

\newpage

\begin{appendix}

\section{PREM Distributional Assumptions and Priors} \label{app:prem_app}

Distributional Assumptions:
\begin{gather*} 
\epsilon_{ji} \overset{\mathrm{iid}}{\sim} N(0,\sigma^2_\epsilon) \\
\\
\boldsymbol{\beta_i} = (\beta_{0i}, \beta_{1i}, \beta_{\{1+1\}i}, ..., \beta_{\{K+1\}i}, \gamma_{1i}, ..., \gamma_{Ki}) \sim MVN(\overline{\boldsymbol{\beta}},\boldsymbol{\Phi_b}) \\
\\
\text{where, } \boldsymbol{\Phi_b} = 
\begin{bmatrix}
\sigma^2_{b_{0}}  \\
0 & \sigma^2_{b_{1}} \\
0 & 0 & \sigma^2_{b_{\{1+1\}}} \\
0 & 0 & 0 & \ddots \\
0 & 0 & 0 & 0 & \sigma^2_{b_{\{K+1\}}} \\
0 & 0 & 0 & 0 & 0 & \sigma^2_{b_{\gamma_{1}}} \\
0 & 0 & 0 & 0 & 0 & 0 & \ddots \\
0 & 0 & 0 & 0 & 0 & 0 & 0 & \sigma^2_{b_{\gamma_{K}}}
\end{bmatrix}
\end{gather*}

Priors:
\begin{itemize}
    \item Error: $\sigma^2_\epsilon \sim \text{Inverse-Gamma}(0.001, 0.001)$
    \item Latent Number of Changepoints $\mathcal{K}$ (2 options):
    \begin{itemize}
        \item $\sim \text{Binomial}(K,p)$ (default; \code{cp\_prior = "binomial"}; modify $p$ using \code{binom\_prob})
        \item $\sim \text{Uniform}(0,...,K)$ (\code{cp\_prior = "uniform"})
    \end{itemize}
    \item Fixed Effects Parameters:
    \begin{itemize}
        \item $\overline{\beta}_0 \sim N(\overline{Y}_{1\cdot}, \text{var}(Y_{1\cdot}))$, where $\overline{Y}_{1\cdot}$ and $\text{var}(Y_{1\cdot})$ are the sample mean and variance of the outcome at the first timepoint.
        \item $\overline{\beta}_k \sim N\Bigl(0, \bigl(\frac{\text{sd}(Y)}{\text{sd}(X)}\bigl)^2\Bigl)$ for $k = \{1, ..., K + 1\}$, where $\text{sd}(X)$ is the sample standard deviation of the timepoints and $\text{sd}(Y)$ is the sample standard deviation of the outcome across all timepoints. 
        \item $\overline{\gamma}_k \sim \text{Uniform}(a,b)$ for $k = \{1, ..., K\}$, where $a$ is the second timepoint and $b$ is the second-to-last timepoint. 
    \end{itemize}
    \item Random Effect Parameters (2 options):
    \begin{itemize}
        \item $\sim \text{Uniform}(0,b)$, where the value of $b$ differs by parameter (default; \code{scale\_prior = “uniform”}).
        \begin{itemize}
            \item $\sigma_{\beta_{0}} \rightarrow b= \text{sd}(Y_{1\cdot})$, where $\text{sd}(Y_{1\cdot})$ is the sample standard deviation of the outcome at the first timepoint. 
            \item $\sigma_{\beta_{k}} \rightarrow b= K\cdot\text{sd}(Y)/\text{sd}(X)$ for $k = \{1, ..., K + 1\}$, where $\text{sd}(X)$ is the sample standard deviation of the timepoints and $\text{sd}(Y)$ is the sample standard deviation of the outcome across all timepoints. 
            \item $\sigma_{\gamma_{k}} \rightarrow b= (\max(X)-\min(X))/4$ for $k = \{1, ..., K\}$
        \end{itemize}
        \item $\sim \text{half-Cauchy}(0,\tau)$	(\code{scale\_prior = “hc”}), where $\tau$ is a parameter-dependent scale parameter that is selected such that the 90th percentile of the distribution equals the upper bound of the default uniform distribution ($b$ above).
    \end{itemize}
\end{itemize}

\newpage

\section{CI-PREM Distributional Assumptions and Priors} \label{app:ciprem_app}

Distributional Assumptions:
\begin{gather*} 
\epsilon_{ji} \overset{\mathrm{iid}}{\sim} N(0,\sigma^2_\epsilon) \\
\\
\boldsymbol{\beta_i} = (\beta_{0i}, \beta_{1i}, \beta_{\{1+1\}i}, ..., \beta_{\{K+1\}i}, \gamma_{1i}, ..., \gamma_{Ki}) \sim MVN(\overline{\boldsymbol{\beta}},\boldsymbol{\Phi_b}) \\
\\
\text{where, } \boldsymbol{\Phi_b} = 
\begin{bmatrix}
\sigma^2_{b_{0}}  \\
0 & \sigma^2_{b_{1}} \\
0 & 0 & \sigma^2_{b_{\{1+1\}}} \\
0 & 0 & 0 & \ddots \\
0 & 0 & 0 & 0 & \sigma^2_{b_{\{K+1\}}} \\
0 & 0 & 0 & 0 & 0 & \sigma^2_{b_{\gamma_{1}}} \\
0 & 0 & 0 & 0 & 0 & 0 & \ddots \\
0 & 0 & 0 & 0 & 0 & 0 & 0 & \sigma^2_{b_{\gamma_{K}}}
\end{bmatrix}
\end{gather*}

Priors:
\begin{itemize}
    \item Error: $\sigma^2_\epsilon \sim \text{Inverse-Gamma}(0.001, 0.001)$
    \item Latent Number of Changepoints $\mathcal{K}$ (2 options):
    \begin{itemize}
        \item $\sim \text{Binomial}(K,p)$ (default; \code{cp\_prior = "binomial"}; modify $p$ using \code{binom\_prob})
        \item $\sim \text{Uniform}(0,...,K)$ (\code{cp\_prior = "uniform"})
    \end{itemize}
    \item Covariates: $\alpha_p \sim N(0,1000)$ for $p=\{1,...,P\}$
    \item Fixed Effects Parameters:
    \begin{itemize}
        \item $\overline{\beta}_0 \sim N(\overline{Y}_{1\cdot}, \text{var}(Y_{1\cdot}))$, where $\overline{Y}_{1\cdot}$ and $\text{var}(Y_{1\cdot})$ are the sample mean and variance of the outcome at the first timepoint.
        \item $\overline{\beta}_k \sim N\Bigl(0, \bigl(\frac{\text{sd}(Y)}{\text{sd}(X)}\bigl)^2\Bigl)$ for $k = \{1, ..., K + 1\}$, where $\text{sd}(X)$ is the sample standard deviation of the timepoints and $\text{sd}(Y)$ is the sample standard deviation of the outcome across all timepoints. 
        \item $\overline{\gamma}_k \sim \text{Uniform}(a,b)$ for $k = \{1, ..., K\}$, where $a$ is the second timepoint and $b$ is the second-to-last timepoint. 
    \end{itemize}
    \item Random Effect Parameters (2 options):
    \begin{itemize}
        \item $\sim \text{Uniform}(0,b)$, where the value of $b$ differs by parameter (default; \code{scale\_prior = “uniform”}).
        \begin{itemize}
            \item $\sigma_{\beta_{0}} \rightarrow b= \text{sd}(Y_{1\cdot})$, where $\text{sd}(Y_{1\cdot})$ is the sample standard deviation of the outcome at the first timepoint. 
            \item $\sigma_{\beta_{k}} \rightarrow b= K\cdot\text{sd}(Y)/\text{sd}(X)$ for $k = \{1, ..., K + 1\}$, where $\text{sd}(X)$ is the sample standard deviation of the timepoints and $\text{sd}(Y)$ is the sample standard deviation of the outcome across all timepoints. 
            \item $\sigma_{\gamma_{k}} \rightarrow b= (\max(X)-\min(X))/4$ for $k = \{1, ..., K\}$
        \end{itemize}
        \item $\sim \text{half-Cauchy}(0,\tau)$	(\code{scale\_prior = “hc”}), where $\tau$ is a parameter-dependent scale parameter that is selected such that the 90th percentile of the distribution equals the upper bound of the default uniform distribution ($b$ above).
    \end{itemize}
\end{itemize}

\newpage

\section{PREMM Distributional Assumptions and Priors} \label{app:premm_app}

Distributional Assumptions:
\begin{gather*} 
\epsilon_{ji} \overset{\mathrm{iid}}{\sim} N(0,\sigma^2_\epsilon) \\
\\
\boldsymbol{\beta_{ci}} = (\beta_{c0i}, \beta_{c1i}, \beta_{c\{1+1\}i}, ..., \beta_{c\{K+1\}i}, \gamma_{c1i}, ..., \gamma_{cKi}) \sim MVN(\overline{\boldsymbol{\beta}}_c,\boldsymbol{\Phi_{b_{c}}})\text{, where } c = \psi(i) \\
\\
\text{and } \boldsymbol{\Phi_{b_{c}}} = 
\begin{bmatrix}
\sigma^2_{b_{c0}}  \\
0 & \sigma^2_{b_{c1}} \\
0 & 0 & \sigma^2_{b_{c\{1+1\}}} \\
0 & 0 & 0 & \ddots \\
0 & 0 & 0 & 0 & \sigma^2_{b_{c\{K+1\}}} \\
0 & 0 & 0 & 0 & 0 & \sigma^2_{b_{\gamma_{c1}}} \\
0 & 0 & 0 & 0 & 0 & 0 & \ddots \\
0 & 0 & 0 & 0 & 0 & 0 & 0 & \sigma^2_{b_{\gamma_{cK}}}
\end{bmatrix}
\end{gather*}

Priors:
\begin{itemize}
    \item Error: $\sigma^2_\epsilon \sim \text{Inverse-Gamma}(0.001, 0.001)$
    \item Latent Number of Changepoints $\mathcal{K}$ (2 options):
    \begin{itemize}
        \item $\sim \text{Binomial}(K,p)$ (default; \code{cp\_prior = "binomial"}; modify $p$ using \code{binom\_prob})
        \item $\sim \text{Uniform}(0,...,K)$ (\code{cp\_prior = "uniform"})
    \end{itemize}
    \item Class Probability: $\pi_{ic} \sim \text{Dirichlet}(\alpha)$ where $\alpha=1$ by default but can be modified using the \code{alpha} argument
    \item Fixed Effects Parameters:
    \begin{itemize}
        \item $\overline{\beta}_{c0} \sim N(\overline{Y}_{1\cdot}, \text{var}(Y_{1\cdot}))$, where $\overline{Y}_{1\cdot}$ and $\text{var}(Y_{1\cdot})$ are the sample mean and variance of the outcome at the first timepoint.
        \item $\overline{\beta}_{ck} \sim N\Bigl(0, \bigl(\frac{\text{sd}(Y)}{\text{sd}(X)}\bigl)^2\Bigl)$ for $k = \{1, ..., K + 1\}$, where $\text{sd}(X)$ is the sample standard deviation of the timepoints and $\text{sd}(Y)$ is the sample standard deviation of the outcome across all timepoints. 
        \item $\overline{\gamma}_{ck} \sim \text{Uniform}(a,b)$ for $k = \{1, ..., K\}$, where $a$ is the second timepoint and $b$ is the second-to-last timepoint. 
    \end{itemize}
    \item Random Effect Parameters (2 options):
    \begin{itemize}
        \item $\sim \text{Uniform}(0,b)$, where the value of $b$ differs by parameter (default; \code{scale\_prior = “uniform”}).
        \begin{itemize}
            \item $\sigma_{\beta_{c0}} \rightarrow b= \text{sd}(Y_{1\cdot})$, where $\text{sd}(Y_{1\cdot})$ is the sample standard deviation of the outcome at the first timepoint. 
            \item $\sigma_{\beta_{ck}} \rightarrow b= K\cdot\text{sd}(Y)/\text{sd}(X)$ for $k = \{1, ..., K + 1\}$, where $\text{sd}(X)$ is the sample standard deviation of the timepoints and $\text{sd}(Y)$ is the sample standard deviation of the outcome across all timepoints. 
            \item $\sigma_{\gamma_{ck}} \rightarrow b= (\max(X)-\min(X))/4$ for $k = \{1, ..., K\}$
        \end{itemize}
        \item $\sim \text{half-Cauchy}(0,\tau)$	(\code{scale\_prior = “hc”}), where $\tau$ is a parameter-dependent scale parameter that is selected such that the 90th percentile of the distribution equals the upper bound of the default uniform distribution ($b$ above).
    \end{itemize}
\end{itemize}

\newpage

\section{CI-PREMM Distributional Assumptions and Priors} \label{app:cipremm_app}

Distributional Assumptions:
\begin{gather*} 
\epsilon_{ji} \overset{\mathrm{iid}}{\sim} N(0,\sigma^2_\epsilon) \\
\\
\boldsymbol{\beta_{ci}} = (\beta_{c0i}, \beta_{c1i}, \beta_{c\{1+1\}i}, ..., \beta_{c\{K+1\}i}, \gamma_{c1i}, ..., \gamma_{cKi}) \sim MVN(\overline{\boldsymbol{\beta}}_c,\boldsymbol{\Phi_{b_{c}}})\text{, where } c = \psi(i) \\
\\
\text{and } \boldsymbol{\Phi_{b_{c}}} = 
\begin{bmatrix}
\sigma^2_{b_{c0}}  \\
0 & \sigma^2_{b_{c1}} \\
0 & 0 & \sigma^2_{b_{c\{1+1\}}} \\
0 & 0 & 0 & \ddots \\
0 & 0 & 0 & 0 & \sigma^2_{b_{c\{K+1\}}} \\
0 & 0 & 0 & 0 & 0 & \sigma^2_{b_{\gamma_{c1}}} \\
0 & 0 & 0 & 0 & 0 & 0 & \ddots \\
0 & 0 & 0 & 0 & 0 & 0 & 0 & \sigma^2_{b_{\gamma_{cK}}}
\end{bmatrix}
\end{gather*}

Priors:
\begin{itemize}
    \item Error: $\sigma^2_\epsilon \sim \text{Inverse-Gamma}(0.001, 0.001)$
    \item Latent Number of Changepoints $\mathcal{K}$ (2 options):
    \begin{itemize}
        \item $\sim \text{Binomial}(K,p)$ (default; \code{cp\_prior = "binomial"}; modify $p$ using \code{binom\_prob})
        \item $\sim \text{Uniform}(0,...,K)$ (\code{cp\_prior = "uniform"})
    \end{itemize}
    \item Covariates:
    \begin{itemize}
        \item $\alpha_p \sim N(0,1000)$ for $p = \{1,...,P\}$
        \item $\lambda_l\sim N(0,10)$ for $l = \{l,...,L\}$
    \end{itemize}
    \item Fixed Effects Parameters:
    \begin{itemize}
        \item $\overline{\beta}_{c0} \sim N(\overline{Y}_{1\cdot}, \text{var}(Y_{1\cdot}))$, where $\overline{Y}_{1\cdot}$ and $\text{var}(Y_{1\cdot})$ are the sample mean and variance of the outcome at the first timepoint.
        \item $\overline{\beta}_{ck} \sim N\Bigl(0, \bigl(\frac{\text{sd}(Y)}{\text{sd}(X)}\bigl)^2\Bigl)$ for $k = \{1, ..., K + 1\}$, where $\text{sd}(X)$ is the sample standard deviation of the timepoints and $\text{sd}(Y)$ is the sample standard deviation of the outcome across all timepoints. 
        \item $\overline{\gamma}_{ck} \sim \text{Uniform}(a,b)$ for $k = \{1, ..., K\}$, where $a$ is the second timepoint and $b$ is the second-to-last timepoint. 
    \end{itemize}
    \item Random Effect Parameters (2 options):
    \begin{itemize}
        \item $\sim \text{Uniform}(0,b)$, where the value of $b$ differs by parameter (default; \code{scale\_prior = “uniform”}).
        \begin{itemize}
            \item $\sigma_{\beta_{c0}} \rightarrow b= \text{sd}(Y_{1\cdot})$, where $\text{sd}(Y_{1\cdot})$ is the sample standard deviation of the outcome at the first timepoint. 
            \item $\sigma_{\beta_{ck}} \rightarrow b= K\cdot\text{sd}(Y)/\text{sd}(X)$ for $k = \{1, ..., K + 1\}$, where $\text{sd}(X)$ is the sample standard deviation of the timepoints and $\text{sd}(Y)$ is the sample standard deviation of the outcome across all timepoints. 
            \item $\sigma_{\gamma_{ck}} \rightarrow b= (\max(X)-\min(X))/4$ for $k = \{1, ..., K\}$
        \end{itemize}
        \item $\sim \text{half-Cauchy}(0,\tau)$	(\code{scale\_prior = “hc”}), where $\tau$ is a parameter-dependent scale parameter that is selected such that the 90th percentile of the distribution equals the upper bound of the default uniform distribution ($b$ above).
    \end{itemize}
\end{itemize}

\newpage

\section{BPREM Distributional Assumptions and Priors} \label{app:bprem_app}

Distributional Assumptions:
\begin{gather*} 
\boldsymbol{\epsilon_{i}} = (\epsilon_{1ji}, \epsilon_{2ji}) \sim MVN(0,\boldsymbol{\Lambda}) \text{, where } \boldsymbol{\Lambda} = 
\begin{bmatrix} 
\sigma^2_{\epsilon_1}  \\
\sigma_{\epsilon_1\epsilon_2} & \sigma^2_{\epsilon_2} \\
\end{bmatrix}
=
\begin{bmatrix} 
\sigma^2_{\epsilon_1}  \\
\rho\sigma_{\epsilon_1}\sigma_{\epsilon_2} & \sigma^2_{\epsilon_2} \\
\end{bmatrix} \\
\\
\boldsymbol{\beta_{i}} = (\beta_{10i}, \beta_{11i}, \beta_{12i},\gamma_{1i},\beta_{20i}, \beta_{21i}, \beta_{22i},\gamma_{2i}) \sim MVN(\overline{\boldsymbol{\beta}},\boldsymbol{\Phi_b}) \\
\\
\text{and } \boldsymbol{\Phi_{b}} = 
\begin{bmatrix}
\sigma^2_{b_{10}}  \\
\sigma_{b_{11}b_{10}} & \sigma^2_{b_{11}} \\
\sigma_{b_{12}b_{10}} & \sigma_{b_{12}b_{11}} & \sigma^2_{b_{12}} \\
\sigma_{b_{\gamma_{1}}b_{10}} & \sigma_{b_{\gamma_{1}}b_{11}} & \sigma_{b_{\gamma_{1}}b_{12}} & \sigma^2_{b_{\gamma_{1}}} \\
\sigma_{b_{20}b_{10}} & \sigma_{b_{20}b_{11}} & \sigma_{b_{20}b_{12}} & \sigma_{b_{20}b_{\gamma_{1}}} & \sigma^2_{b_{20}} \\
\sigma_{b_{21}b_{10}} & \sigma_{b_{21}b_{11}} & \sigma_{b_{21}b_{12}} & \sigma_{b_{21}b_{\gamma_{1}}} & \sigma_{b_{21}b_{20}} & \sigma^2_{b_{21}} \\
\sigma_{b_{22}b_{10}} & \sigma_{b_{22}b_{11}} & \sigma_{b_{22}b_{12}} & \sigma_{b_{22}b_{\gamma_{1}}} & \sigma_{b_{22}b_{20}} & \sigma_{b_{22}b_{21}} & \sigma^2_{b_{22}} \\
\sigma_{b_{\gamma_{2}}b_{10}} & \sigma_{b_{\gamma_{2}}b_{11}} & \sigma_{b_{\gamma_{2}}b_{12}} & \sigma_{b_{\gamma_{2}}b_{\gamma_{1}}} & \sigma_{b_{\gamma_{2}}b_{20}} & \sigma_{b_{\gamma_{2}}b_{21}} & \sigma_{b_{\gamma_{2}}b_{22}} & \sigma^2_{b_{\gamma_{2}}}
\end{bmatrix}
\end{gather*}

Priors:
\begin{itemize}
    \item Error: 
    \begin{itemize}
        \item $\sigma_{\epsilon_{k}} \sim \text{Uniform}(0,\min(\text{var}(Y_{kj}))$ for $k=1,2$, where $\text{var}(Y_{kj})$ is the sample variance of outcome variable $k$ at measurement occasion $j$
        \item $\rho \sim \text{Uniform}(-1,1)$
    \end{itemize}
    \item Fixed Effects Parameters:
    \begin{itemize}
        \item $\overline{\beta}_{k0}, \overline{\beta}_{k1}, \overline{\beta}_{k2} \sim N(0,10000)$ for $k=1,2$
        \item $\overline{\gamma}_{k} \sim N(a,b^2)T(\min(X),\max(X))$ for $k = 1,2$, where $a=(\max(X)-\min(X))/2$ and $b=(\max(X)-\min(X))/4$. 
    \end{itemize}
    \item Random Effect Parameters: $\boldsymbol{\Phi_{b}} \sim \text{scaled inverse-Wishart}$ \citep{Peralta2022Bayesian}
\end{itemize}

\newpage

\section{CREM Distributional Assumptions and Priors} \label{app:crem_app}

Distributional Assumptions:
\begin{equation} 
\epsilon_{jir} \overset{\mathrm{iid}}{\sim} N(0,\sigma^2_\epsilon) \\
\end{equation}

For the CREM, the random coefficients for each parameter $q$ can be broken down as follows: 

\begin{equation} 
\beta_{qir} = \beta_q + b_{qi} + g_{qr}
\end{equation}

where $\beta_q$ is the fixed effect and 

\begin{gather*}
\boldsymbol{b_{i}} \sim MVN(\boldsymbol{0},\boldsymbol{\Phi_b}) \qquad
\boldsymbol{\Phi_b} = 
\begin{bmatrix}
\sigma^2_{b_{0}}  \\
\vdots & \ddots \\
\sigma_{b_{q}b_{0}} & \dots & \sigma^2_{b_{q}} \\
\end{bmatrix}\\
\boldsymbol{g_{r}} \sim MVN(\boldsymbol{0},\boldsymbol{\Phi_g}) \qquad
\boldsymbol{\Phi_g} = 
\begin{bmatrix}
\sigma^2_{g_{0}}  \\
\vdots & \ddots \\
\sigma_{g_{q}g_{0}} & \dots & \sigma^2_{g_{q}} \\
\end{bmatrix}
\end{gather*}

Priors:
\begin{itemize}
    \item Error: $\sigma^2_\epsilon \sim \text{Inverse-Gamma}(0.001, 0.001)$
    \item Fixed Effects Parameters:
    \begin{itemize}
        \item $\overline{\beta}_{q} \sim N(0,100000)$ for $k=1,2$
        \item $\overline{\gamma} \sim \text{Uniform}(a,b)$ for $k = \{1, ..., K\}$, $a$ is the second timepoint and $b$ is the second-to-last timepoint (applies to \code{form = "piecewise"} only)
    \end{itemize}
    \item Random Effect Parameters: $\boldsymbol{\Phi_{b}} \sim \text{scaled inverse-Wishart}$ \citep{Peralta2022Bayesian}
\end{itemize}

\newpage

\begin{landscape}

\section{Extraction Methods Available in BEND} \label{app:extract_methods}

The below table provides an overview of the extraction methods available in \pkg{BEND} and comparable methods from other packages (if it exists). 

\begin{table}[h]
\begin{tabular}{@{}llll@{}}
\toprule
\thead{Extraction\\Method} & \thead{Definition}                                                & \thead{Comparable\\Methods} & \thead{Notes}                                                                                                                                                              \\ \midrule
getFixEf()             & Fixed effects estimates                                   & fixef()                                            &                                                                                                                                                                    \\
getRanEf()             & Random effects estimates                                  & ranef()                                            & \thead[l]{Only available for Bayes\_CREM and Bayes\_BPREM. \\In Bayes\_PREM, random coefficients are estimated directly \\(thus, random effects are not included in the output).} \\
getCoef()              & Random coefficient estimates                              & coef()                                             &                                                                                                                                                                    \\
getVarCov()            & Random effects variance-covariance matrix                 & getVarCov()                  & \thead[l]{Bayes\_PREM does not estimate covariances, so the off-diagonal\\will be blank. }                                                                                     \\
getFitted()            & Fitted values                                             & fitted()                                           &                                                                                                                                                                    \\
getModelFit()          & Model fit information (deviance, pD, DIC)                 & logLik()                                           &                                                                                                                                                                    \\
getClassProb()         & Class Probabilities                                       &                                                    & for PREM only                                                                                                                                                      \\
getKProb()             & $\mathcal{K}$ probabilities (for each class) &                                                    & for PREM only                                                                                                                                                      \\ \bottomrule
\end{tabular}
\caption{BEND Extraction Methods}
\label{tab:ext_meth}
\end{table}
\end{landscape}

\end{appendix}

%% -----------------------------------------------------------------------------

\end{document}